\documentclass[aps,pre,twocolumn,superscriptaddress,floats,floatfix,showpacs]{revtex4-2}
\usepackage{amssymb,amsmath}
\usepackage{float}
\usepackage{mathrsfs}
\usepackage{graphics}
\usepackage{epstopdf}
\usepackage{color}
\usepackage{subfigure}
\usepackage{epsfig}
\usepackage{rotating}
\usepackage{anyfontsize}
\usepackage{verbatim}
\usepackage{standalone}
\usepackage{hyperref}
\usepackage{array}

\usepackage{xr}

\usepackage{graphicx}
\usepackage{dcolumn}
\usepackage{bm}
\usepackage{xcolor}
\usepackage{amsmath,amsfonts,amssymb,amsthm}
\usepackage{mathtools}
\usepackage{commath}
\usepackage{color,soul}
\begin{document}

\title{Spiral-wave dynamics 
in excitable media: Insights from dynamic mode decomposition} 


\author{Mahesh Kumar Mulimani}
\email{maheshk@iisc.ac.in ;}
\affiliation{Centre for Condensed Matter Theory, Department of Physics, Indian Institute of Science, Bangalore 560012, India.}
\author{Soling Zimik}%
\email{solyzk@gmail.com ;}
\affiliation{Computational Biology Group, Institute of Mathematical Sciences, CIT Campus, Tharamani, Chennai, 600113, India.}
\author{Jaya Kumar Alageshan}
\email{jayak@iisc.ac.in ;}
\affiliation{Centre for Condensed Matter Theory, Department of Physics, Indian Institute of Science, Bangalore 560012, India.}
\author{Rahul Pandit}
 \email{rahul@iisc.ac.in.}
 \affiliation{Centre for Condensed Matter Theory, Department of Physics, Indian Institute of Science, Bangalore 560012, India.}

\begin{abstract}

Spiral waves are ubiquitous spatiotemporal patterns that occur in various
excitable systems. In cardiac tissue, the formation of these spiral
waves is associated with life-threatening arrhythmias, and, therefore,
it is important to study the dynamics of these waves. Tracking the
trajectory of a spiral-wave tip can reveal important dynamical features
of a spiral wave, such as its periodicity, and its vulnerability to
instabilities.  We show how to employ the data-driven
spectral-decomposition method, called dynamic mode  decomposition
(DMD), to detect a spiral tip trajectory (TT) in three settings: (1) a
homogeneous medium; (2) a heterogeneous medium; and (3) with external
noise. We demonstrate that the performance of DMD-based TT (DMDTT) is
either comparable to or better than the conventional tip-tracking
method, called the isopotential-intersection method (IIM), in the 
cases (1)-(3): (1) Both IIM and DMDTT
capture TT patterns at small values of the image-sampling interval
$\tau$; however, IIM is more sensitive than DMDTT to the changes in $\tau$.
(2) In a heterogeneous medium, IIM yields TT patterns, but with a background of 
scattered noisy points, which are suppressed in DMDTT. 
(3) DMDTT is more robust to external noise than IIM.
We show, finally, that DMD can be used to reconstruct, and
hence predict, the spatiotemporal evolution of spiral waves in the
models we study.  

\end{abstract}

\keywords{spiral waves; tip trajectories; excitable media; cardiac tissue;
dynamic mode decomposition }
\maketitle


\section{Introduction}
\label{sec:Introduction}

The spatiotemporal organization of nonlinear waves into spirals occurs in
various excitable
media~\cite{winfree1972spiral,lechleiter1991spiral,tyson1989cyclic,falcke1992traveling,seiden2015tongue,mcguire2021geographic,marino2019excitable}.
In cardiac tissue, the formation of such spiral waves of electrical activation 
is associated with life-threatening cardiac
arrhythmias~\cite{allessie1977circus,davidenko1990sustained,pertsov1993spiral,gray1995mechanisms,gray1998spatial}.
A stable rotating spiral wave is linked to ventricular tachycardia (VT), i.e.,
rapid heart beats; a spiral wave,  with a meandering  core, can cause polymorphic
VT (with aperiodic heart beats); and a multiple-spiral state is linked to
ventricular fibrillation (VF) and chaotic heart beats. These arrhythmias are a
leading cause of death. It is, therefore, crucial to understand the dynamics of
a spiral wave in cardiac tissue.  By tracking the trajectory of the phase
singularity at the tip of a spiral wave, we can reveal some of its dynamical
features. For example, the tip of a stably rotating spiral wave, with one
fundamental frequency, typically traces a circular trajectory; a spiral wave
that rotates with two or more incommensurate fundamental frequencies can
exhibit complicated tip-trajectory (TT)
patterns~\cite{fenton1998vortex,qu2000origins,gray2009origin}, which can lead
to irregular heart rates. Such complicated TTs are prone to
instabilities~\cite{qu2000origins} that cause the spiral wave to break up into
multiple daughter spiral waves, which lead in turn to a spiral-turbulent state.
Therefore, tracking the TT of a spiral wave yields valuable insights into its
dynamics; and the detection of phase singularities of spiral waves, in
\textit{ex-vivo} and \textit{in-vivo} experiments and in \textit{in silico}
studies of mathematical models, is of great
importance~\cite{aronis2017rotors,umapathy2010phase,bray2001experimental,gray1998spatial}.
In particular, the location of such phase singularities, with high specificity
and selectivity, can be employed for accurate ablation, which can help in 
the termination of life-threatening arrhythmias.  

We show that the data-driven spectral-decomposition method known as
\textit{dynamic mode decomposition} (DMD), which has been used to analyze
complex spatiotemporal evolution in a variety of spatially extended
systems~\cite{Schmid2010,Schmid2011,Schmid2011b,nathan2018applied,DMDBook,krake2021visualization,tu2013dynamic},
can be used fruitfully in excitable medial (a) to identify spiral-wave TTs and
(b) to study the spatiotemporal evolution of spiral waves.  Although the DMD
method has been used to uncover \textit{coherent structures} in fluid flows, to
the best of our knowledge it has not been used to study nonlinear waves in
mathematical models for cardiac tissue.  \textcolor{black}{Two studies
have applied DMD to spiral waves: one that deals with the extraction of an
approximate governing equation for the spiral waves~\cite{champion2019data};
and the other study that extracts observables that
are possible candidates for Koopman operators~\cite{nathan2018applied}. 	
Our application of DMD to spiral waves in mathematical models for 
excitable media and cardiac tissue leads to new insights
into spiral-tip trajectories and the prediction of the dynamics of
these waves.}

We use DMD to investigate the spatiotemporal evolution of spiral waves in two
mathematical models for cardiac tissue; and we show that DMD can be used
effectively (a) to identify TT patterns and (b) to reconstruct and predict the
spatiotemporal evolution of spiral waves by using the DMD eigenmodes.  There
are conventional methods of tracking TT in \textit{in-silico} studies, among
which the most common is the isopotential-intersection method
(IIM)~\cite{fenton1998vortex}.  We compare the versatility of the 
DMDTT method relative to the IIM technique in three different settings: 
(1) a homogeneous medium;
(2) a heterogeneous medium; and (3) with external noise. We compare the
performance of DMD-based TT (DMDTT) with the conventional isopotential
intersection method (IIM)~\cite{fenton1998vortex} and show that
the former is either comparable to or better and more versatile than the
latter: In case (1), both IIM and DMDTT capture TT patterns at small values of
the image-sampling interval $\tau$; however, IIM is more sensitive than DMDTT
to changes in $\tau$.  In case (2), we find that IIM yields TT patterns,
but with a background of scattered noisy points; by contrast, DMDTT does not
lead to such noise.  In case (3), we show that DMDTT is more robust to external
noise than IIM.  We show, finally, that DMD can be used to reconstruct, and
hence predict, the spatiotemporal evolution of spiral waves in the models we
study. 

\begin{figure}[!ht]
    \centering
    \includegraphics[scale=0.188]{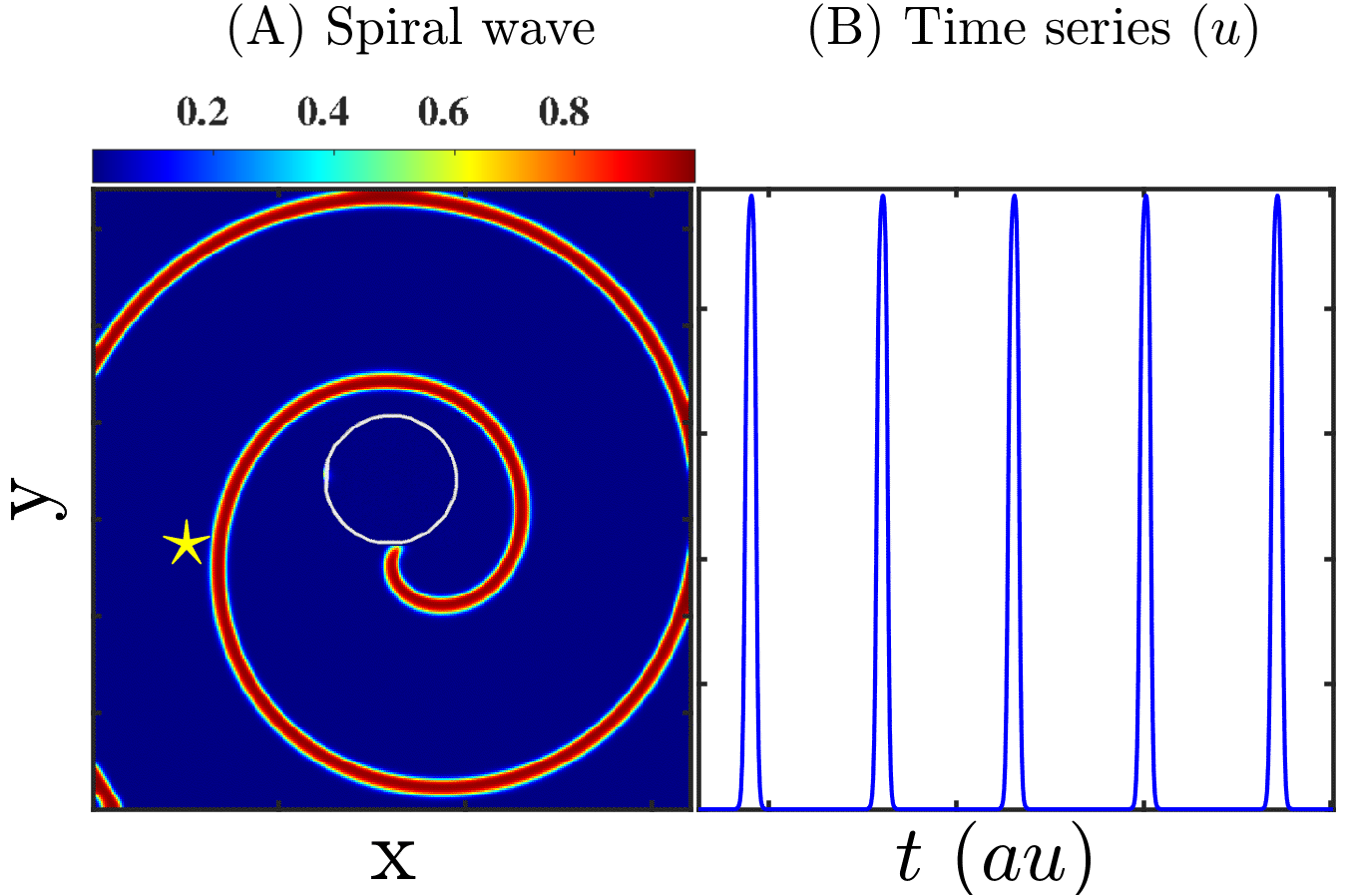}

\caption{(A) Pseudocolor plot of $u$, for the Barkley model [Eqs.~(\ref{B1})
and (\ref{B2})], showing a spiral wave, whose tip traces a circular
trajectory (white curve). The yellow star indicates the point at which
we record the time series of $u$ to obtain the frequency of the spiral
waves; (B) a plot of the time series of $u$ recorded from the point
indicated by the yellow star; such plots yield the frequency of the
spiral wave.}

\label{fig:fig0}
\end{figure}

\begin{figure*}[!ht]
   \centering
   \includegraphics[scale=0.18]{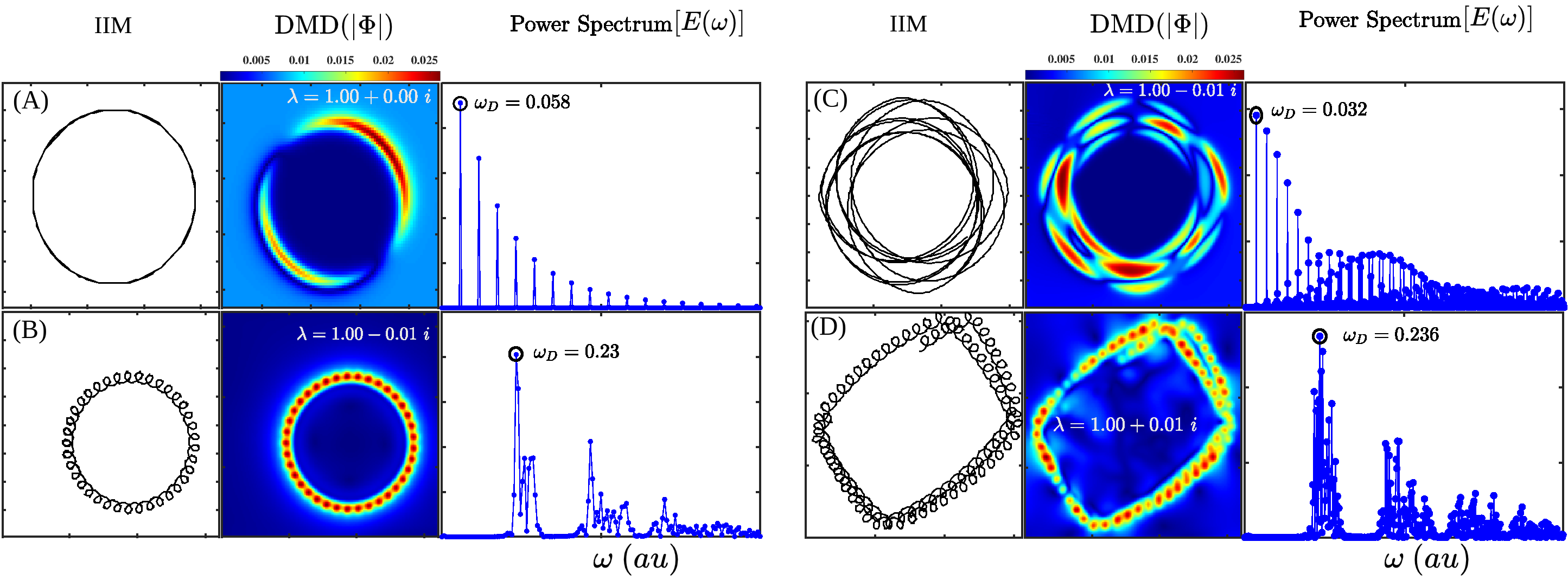}
\caption{Illustrative plots, for the Barkley model [Eqs.~(\ref{B1})
and (\ref{B2})], showing four different types of tip-trajectories
(TTs). In each one of the panels (A)-(D), the sub-figures in the left
columns show the TTs obtained from the IIM (see text), those in the
middle columns show, pseudocolor plots, the modulus of 
a DMD eigenmode $\Phi$ (see text), with eigenvalue $\lambda$,
that contains an imprint of the TT, and those in the right column show
the power spectrum, obtained from the time series of $u$ (cf.
Fig.~\ref{fig:fig0}).  The circular TT in (A) arises from periodic
motion, with a fundamental frequency $\omega_D=0.05$, which appears
clearly in the power spectrum along with its higher harmonics. The TTs
in (B)-(D) arise from aperiodic motions with more than one fundamental
frequency (see text).}
\label{fig:fig1}
\end{figure*}


The remainder of our paper is organized as follows. In Section~\ref{sec:MM}, on
Models and Methods, we describe the mathematical models, numerical schemes,
DMD, and the tip tracking IIM that  we use in our study. We present the findings of our study in
Section~\ref{sec:Results}.  Finally, in Section~\ref{sec:Conclusions}, we
summarize our conclusions and provide a discussion of our results in the light
of earlier studies.

\section{Models and Methods}
\label{sec:MM}

We begin with a description of the mathematical models we use in
Subsection~\ref{subsec:Models}.  In Subsection~\ref{subsec:IIM} we present a
brief overview of the conventional IIM.  We give a short introduction to the
DMD methods we use in Subsection~\ref{subsec:DMD}.

\subsection{Reaction-diffusion models for electrical-excitation waves
in cardiac tissue}
\label{subsec:Models}

We use two mathematical models for cardiac tissue to illustrate
the application of DMD to the study of spiral waves in excitable media.
The first is the set of two-variable coupled partial differential equations
(PDEs) called the Barkley model~\cite{Barkley1991}; and 
the second is the biophysically realistic O'Hara-Rudy (ORd) 
model~\cite{o2011simulation}. The Barkley-model PDEs are:
\begin{eqnarray}
\label{B1} 
\frac{\partial{u}}{\partial{t}} & = & \frac{1}{\epsilon}u(1-u)(u-\frac{v+b}{a})+D\nabla^2{u}; \\	
\frac{dv}{dt}  &=&  u-v .
\label{B2}
\end{eqnarray} 
$u$ and $v$ are the fast-excitation and slow-recovery variables, respectively,
at the point $\bf{r}$ and time $t$; the time-scale separation between $u$ and
$v$ is controlled by the value of $\epsilon$; the parameter $a$ sets the
duration of excitation, and $\frac{b}{a}$ sets the threshold of excitation. $D$
is the diffusion constant of the medium (we use $D=1$). We solve
Eqs.~(\ref{B1}) and (~\ref{B2}) by using the forward-Euler method for time
marching and a five-point stencil for the Laplacian. The temporal and spatial
resolutions are set to be $\Delta x = 0.2$ (arbitrary units $au$)
and $\Delta t=0.005$ $au$. 

For our study with inexcitable obstacles in the medium, we use the  O'Hara-Rudy
(ORd) model~\cite{o2011simulation} for cardiac myocytes. In a homogeneous
medium, the ORd model uses the following PDE for the transmembrane potemtial
$V_m(\bf{r},t)$:

\begin{eqnarray}
        \frac{\partial{V_m}}{\partial{t}} &=& D\ {\nabla^2}V_m - \frac{I_{ion}}{C_{m}} ; 
\nonumber \\
I_{\rm{ion}} &=& I_{Na} + I_{to} + I_{CaL} + I_{CaNa} + I_{CaK} + I_{Kr} + 
I_{Ks}  \nonumber \\
&+& I_{K1} + I_{NaCa} + I_{NaK} + I_{Nab} + I_{Cab} \nonumber \\
&+& I_{Kb} + I_{pCa} ;
\label{eq:VmPDE}
\end{eqnarray}
here, $C_{m}$ is the membrane capacitance, $D$ the diffusion coefficient (for
simplicity, chosen to be a scalar), and the total ionic current $I_{\rm{ion}}$
is a sum of the following ion-channel currents: the fast inward Na$^+$ current
$I_{Na}$; the transient outward K$^+$ current $I_{to}$ ; the L-type Ca$^{2+}$
current $I_{CaL}$; the Na$^+$ current through the L-type Ca$^{2+}$ channel
$I_{CaNa}$ ; the $K^+$ current through the L-type Ca$^{2+}$ channel $I_{CaK}$ ;
the rapid delayed rectifier $K^+$ current $I_{Kr}$; the slow delayed rectifier
$K^+$ current $I_{Ks}$; the inward rectifier $K^+$ current $I_{K1}$ ; the
Na$^+/$Ca$^{2+}$ exchange current $I_{NaCa}$; the Na$^+/$K$^+$ ATPase current
$I_{NaK}$; the Na$^+$ background current $I_{Nab}$; the Ca$^{2+}$ background
current $I_{Cab}$; the K$^+$ background current $I_{Kb}$; the sarcolemmal
Ca$^{2+}$ pump current $I_{pCa}$; for a full list of these currents and the
equations that govern their evolution we refer the reader to
Refs.~\cite{o2011simulation,zimik2015computational,zimik2017reentry}, which
also describe the finite-difference numerical methods that we use.  In both
the models we study, we restrict ourselves to two spatial dimensions and we
employ no-flux boundary conditions.

Excitation waves in the Barkley model [Eqs.~(\ref{B1}) and (~\ref{B2})] have
small wavelengths, so they are vulnerable to wavebreaks, especially in the
presence of heterogeneities in the medium. In contrast, the ORd model
[Eq.~(\ref{eq:VmPDE})] yields waves with large wavelengths; these waves are
more stable in heterogeneous media than their Barkley-model counterparts. As
our objective is to investigate the detection of the phase singularity of a
stable spiral wave, we use the ORd model for our study with a heterogeneous
medium; the coupling between cells in the presence of inexcitable obstacles is
modelled as in Refs.~\cite{ten2007influence,zimik2017reentry}.

\subsection{Isopotential intersection method (IIM)}
\label{subsec:IIM}

The tracking of the tip of a spiral wave, as described 
in Ref.~\cite{fenton1998vortex}, is based
on the idea that the normal velocity of the spiral wave at its tip is zero.
We illustrate this for the Barkley model Eqs.~\ref{B1}-\ref{B2}, in which
$du/dt=0$ at the spiral tip. We choose isopotential lines
with value $u_{iso}=0.4-0.5$; we then track the intersection point of 
$u(x,y)$ with $u_{iso}$ at different times, i.e., we record the 
position $(x,y)$, at a given time, where.
\begin{equation}
u(x,y)-u_{iso}=0.
\end{equation}
The locus of these intersection points gives the spiral TT.

\subsection{Dynamic Mode Decomposition (DMD)}
\label{subsec:DMD}

The data-driven DMD method employs a linear operator to model the
spatiotemporal evolution of fields in a complex, typically nonlinear, system.
If $\mathbf{x}_n$ represents the vector form of some spatial data (typically an
image) at the $n^{th}$ time instant, then the best estimate of the linear
operator $\mathcal{L}$ that translates $\mathbf{x}_n$ to its value
$\mathbf{x}_{n+1}$, at the next time step, follows from the minimization problem

\begin{eqnarray}
    \min_\mathcal{L} \left\{ \sum_{n=o}^{m-1} \norm{ \mathcal{L} \; \mathbf{x}_n - \mathbf{x}_{n+1} }_2 \right\} ,
    \label{Norm}
\end{eqnarray}

where $\norm{ . }_2$ represents the $L^2$-norm, and ${m}$ is the total number
of images (spatial data), each one of which is separated from its predecessor 
and successor by the sampling-time interval $\tau$. The eigenvalues and
eigenvectors of $\mathcal{L}$ contain useful information about the evolution of
the system and can be calculated efficiently by using singular value
decomposition~\cite{darling2019eigenfunctions,DMDBook}. In our analysis we
choose ${m}$ such that it is greater than all the macroscopic time-scales
(spiral-rotational periods) that are present in our system.
We construct a matrix $X_1$, with column vectors of images at discrete
times labelled $0, 1, \ldots, (m-1)$, and a similar matrix $X_2$, with column 
vectors of images at discrete times labelled $1, 2, \ldots, m$ as follows:

\begin{eqnarray}
    X_1=\begin{bmatrix}
        | &  & |\\
        x_0 & ... & x_{m-1}\\
        |& &|
        \end{bmatrix};
    X_2=\begin{bmatrix}
        | &  & |\\
        x_1 & ... & x_{m}\\
        |& &|
        \end{bmatrix}
	\label{eq:X1X2}.
\end{eqnarray}

The operator that best fits Eq.~(\ref{Norm}) is 
\begin{eqnarray}
    \mathcal{L}=X_2 X_1^{\dagger},
\end{eqnarray}
where $X_1^{\dagger}$ is the Moore-Penrose
pseudoinverse~\cite{Barata2012,penrose1955generalized}. By using the DMD
algorithm~\cite{DMDBook}, we get the following spectral decomposition:
\begin{eqnarray}
    \mathcal{L} \Phi_i=\lambda_i \ \Phi_i ;
\end{eqnarray}
here, $\lambda_i$ and $\Phi_i$ denote the eigenvalue and eigenmode of
$\mathcal{L}$, respectively. Depending on whether $|\lambda_i|$ $>1$,
$|\lambda_i|$ $=1$, or $|\lambda_i|$ $<1$, the corresponding eigenmode $\Phi_i$
grows, remains constant, or decays, respectively, in time.  In Section I of the
Supplemental Material~\cite{supmat} we give the SVD-based method that we use to
get a low-rank version of $\mathcal{L}$ and thence the dominant eigenmodes. We
refer the reader to~\cite{DMDBook} for a detailed discussion of DMD.



\section{Results}
\label{sec:Results}

We begin with our results for TT via DMD in Subsection~\ref{subsec:spiralTT}.
In Subsection~\ref{subsec:heteroTT} we present our DMDTT results in a
heterogeneous-cardiac-tissue model and also in the presence of external noise.
We then give a short introduction to how DMD can be used to reconstruct
spiral-wave dynamics in Subsection~\ref{subsec:reconstruct}.

\begin{figure}
   \centering
       \includegraphics[scale=0.130]{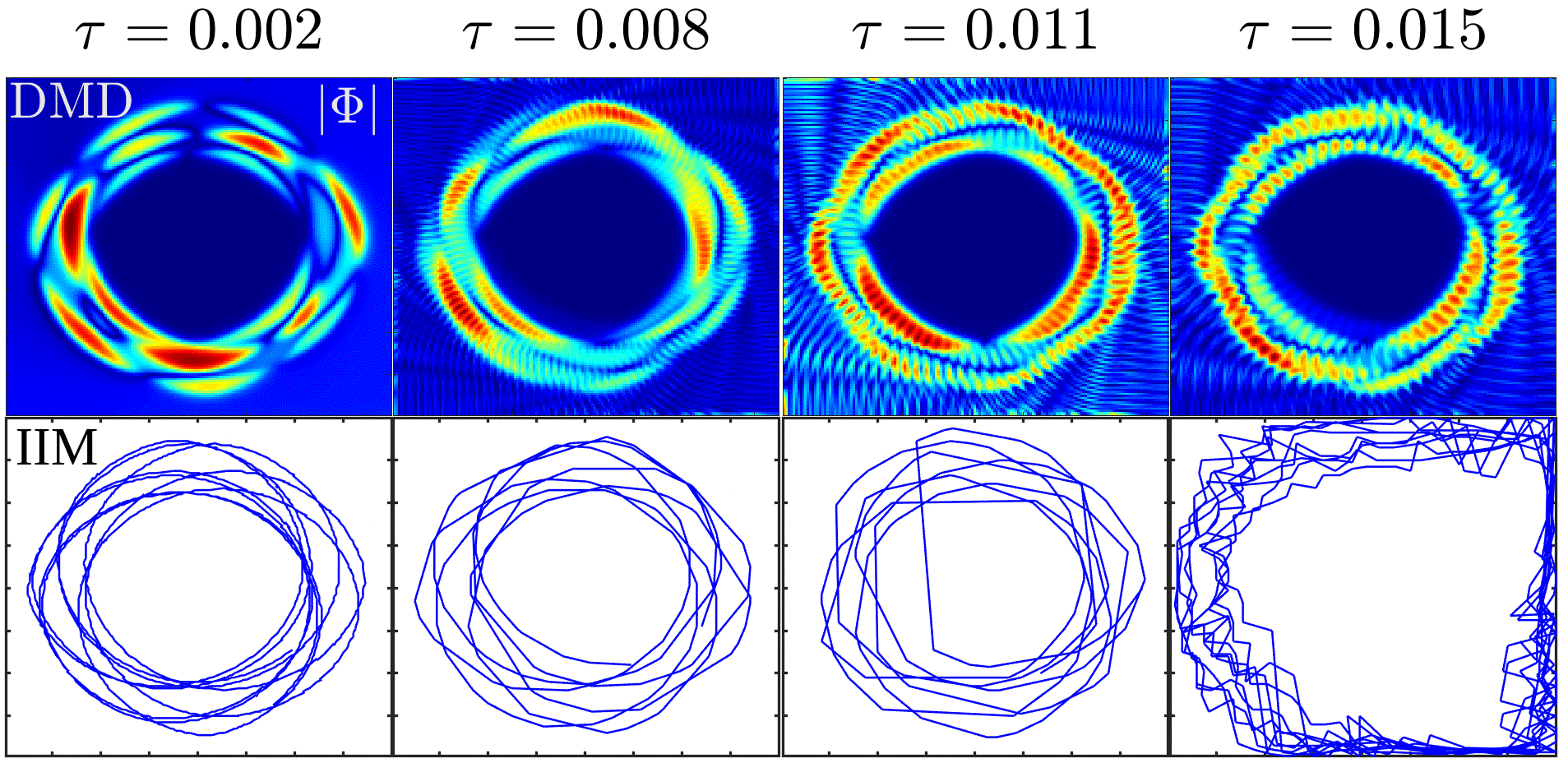}

\caption{Illustrative TT patterns: (Top row) Pseudocolor plots of the
modulus of a DMD eigenmode $\Phi$, with eigenvalue $\lambda$, display  
high intensity along the
tip trajectory. (Bottom row) TTs obtained via the conventional IIM (see
text). We use four different values of the non-dimensionalised sampling
interval $\tau$ (see text) and the rosette pattern of
Fig.~\ref{fig:fig0} (C) for the Barkley model [Eqs.~(\ref{B1}) and
(~\ref{B2})]; IIM TTs are  more sensitive to changes in the value of
$\tau$ as compared to their DMD counterparts.} 

	\label{fig:fig2}
\end{figure}

\begin{figure}
    \centering
    \includegraphics[scale=0.36]{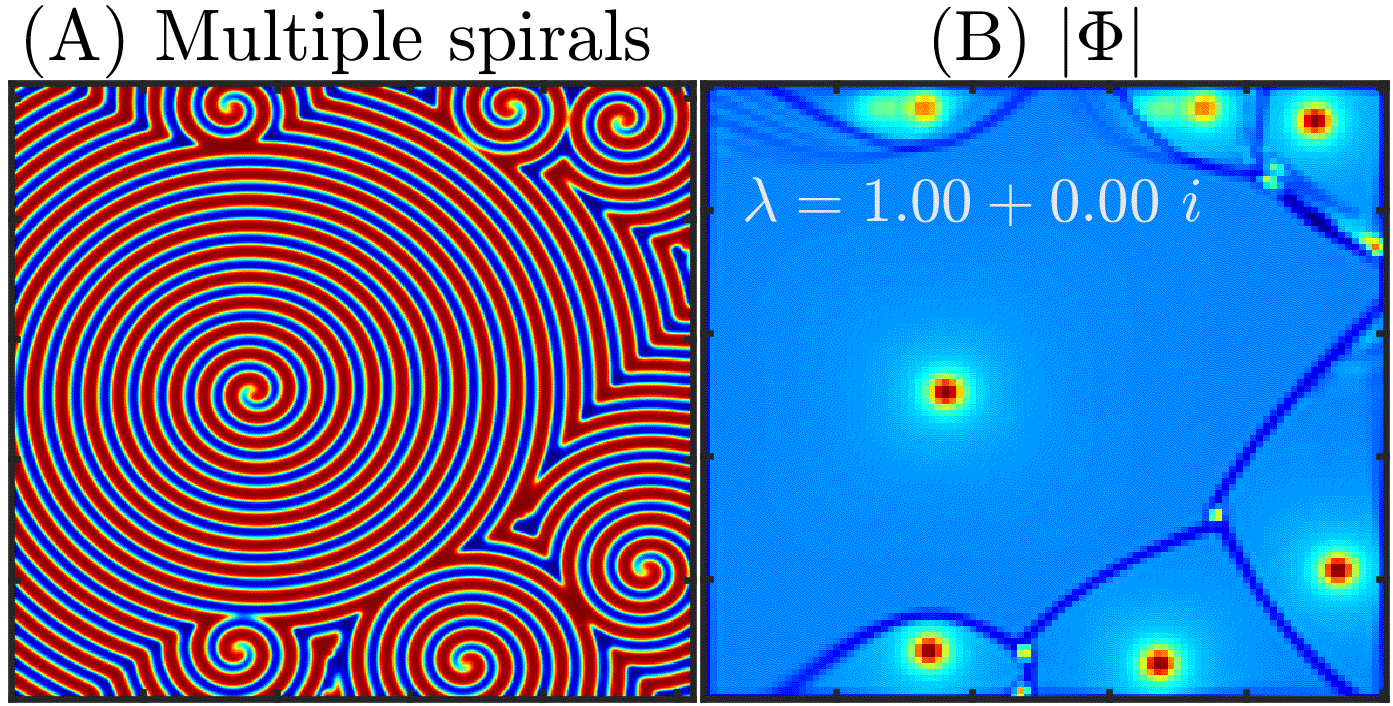}

\caption{Pseudocolor plots of (A) $u$, for the Barkley model [Eqs.~(\ref{B1})
and (~\ref{B2})], showing a multiple-spiral-wave state, and (B) the
modulus of one of the DMD eigenmodes, showing the location of the phase
singularities in this multiple-spiral-wave state and also the
boundaries between the domains in this state.}

\label{fig:fig3} 
\end{figure}

\subsection{Spiral TT in a homogeneous medium}
\label{subsec:spiralTT}

We characterize spiral-wave dynamics here by spiral-TT patterns and the
wave-rotation frequencies. We illustrate this for the Barkley model
[Eqs.~(\ref{B1}) and (~\ref{B2})]. To obtain the frequencies of a spiral wave,
we record the time series of $u$, Fig.~\ref{fig:fig0} (B), from a
representative point, which is marked by a yellow star in the simulation domain
[Fig.~\ref{fig:fig0} (A)]. From the power spectrum $E(\omega)$ of this time
series we obtain the most important frequencies, like the frequency $\omega_D$
of the highest peak in this spectrum. We then use the IIM and DMD methods to
obtain the TTs.  In the illustrative plots of Fig.~\ref{fig:fig1} we show four
different types of TTs, in each one of the panels (A)-(D); the sub-figures in
the left columns show the TTs obtained from the IIM; those in the middle columns
show, via pseudocolor plots of the modulus of a DMD eigenmode $\Phi$, with eigen value $|\lambda|=1$,
that contains an imprint of the TT; the right columns show the power spectra
$E(\omega)$.  The circular TT in Fig.~\ref{fig:fig1} (A) arises from periodic
motion, with a fundamental frequency $\omega_D=0.05$, which appears clearly in
the power spectrum along with its harmonics.  The TTs in Figs.~\ref{fig:fig1}
(B), (C), and (D) exhibit meandering patterns with petals on a circular
trajectory, rosettes, and a rectangular trajectory with petals, respectively;
these patterns arise from aperiodic motions with more than one fundamental
frequency; e.g., the peaks in the $E(\omega)$ in Fig.~\ref{fig:fig1} (B) can be
labelled as $n_1 \omega_D + n_2 \omega_2$, with $n_1$ and $n_2$ integers
(positive or negative), $\omega_D = 0.253333$ and $\omega_2 = 0.245926$, and 
the ratio $\omega_2/\omega_D \simeq 0.9707$ an irrational number. We
show other TT patterns, for different parameter sets, in Fig.  S2 in the
Supplementary Material~\cite{supmat}.  From Fig.~\ref{fig:fig1} we conclude
that both IIM and DMD methods can detect complicated TT patterns equally well
in a homogeneous medium.

%
\begin{table}[!ht]
\resizebox{7cm}{!}{
\begin{tabular}{|l|l|}
\hline
\ \hspace{0.25cm}$\tau = \mathcal{T} \times \Delta t \times \omega_D$  & \ \  \hspace{0.05cm} $\mathcal{T}$ \ \\
\hline
\hline
 \hspace{0.75cm}0.002 \ & \ \hspace{0.15cm}10   \\ \hline

 \hspace{0.75cm}0.008 \ & \ \hspace{0.15cm}50  \\\hline

 \hspace{0.75cm}0.011 \ & \ \hspace{0.15cm}70   \\\hline

  \hspace{0.75cm}0.015 \ & \ \hspace{0.15cm}90   \\
\hline
\end{tabular}}
\caption{The values of the sampling time $\tau$ that are used in Fig.~\ref{fig:fig2} for the rosette TT (Fig.~\ref{fig:fig1}(C)); $\mathcal{T}$ is the number of iterations between successive images (e.g., $x_0$ and $x_1$); we obtain $\omega_D$ from            
Fig.~\ref{fig:fig1}(C); $\Delta t$ (arbitrary units $au$) is the time step in  
our numerical simulations.}
\label{Table:Table1}
\end{table}

We show, in Fig.~\ref{fig:fig2}, how the TT patterns vary with the
non-dimensionalised sampling interval $\tau = \mathcal{T} \times \Delta t
\times \omega_D$ (Table~\ref{Table:Table1}) in both the IIM and DMD methods.
For specificity, we use the rosette pattern in Fig.~\ref{fig:fig2} (C).  The TT
pattern is roughly conserved as we increase $\tau$; however, the IIM yields
noisy  TTs for $\tau \geq 0.015$, whereas the DMD method yields a TT pattern
that is less noisy than its IIM counterpart. 

We now demonstrate that DMD can be used to determine the positions of phase
singularities even when there are multiple spiral waves in the medium. For the
Barkley model [Eqs.~(\ref{B1}) and (~\ref{B2})], we illustrate this in
Fig.~\ref{fig:fig3} (A), which  shows a pseudocolor plot of $u$ in a
multiple-spiral state; and Fig.~\ref{fig:fig3} (B) depicts
the modulus of a DMD eigenmode $\Phi$ that
exhibits clearly the locations of the phase singularities that are present in
this pseudocolor plot. In Fig.~\ref{fig:fig3} (B) we also see the boundaries
between the domains with spiral waves.

\begin{figure}
    \centering
        \includegraphics[scale=0.22]{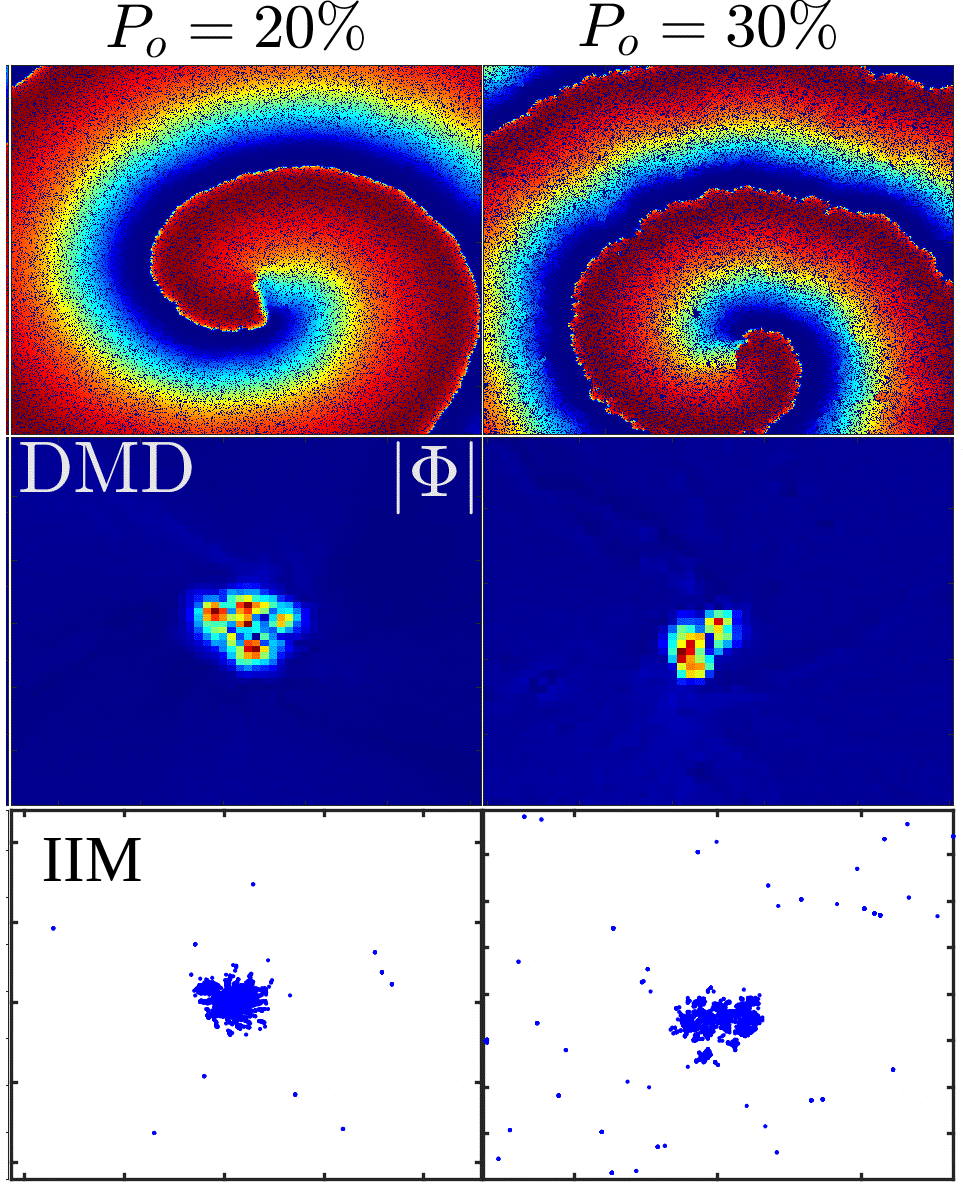}

\caption{The detection of TT patterns, via IIM and DMD, with $P_o = 20\%$ (left
panel) and $P_o = 30\%$ (right panel) heterogeneities (inexcitable
obstacles) in the the ORd model [Eq.~(\ref{eq:VmPDE})]: (Top row)
Pseudocolor plots of $V_m$ showing spiral waves; black dots indicate
inexcitable obstacles. (Middle row) Pseudocolor plots of the modulus of
a DMD eigenmode $\Phi$ display  high intensity along the tip trajectory.
(Bottom row) TTs tracked by the IIM; in addition to the TT, this IIM
shows randomly scattered points, which do not appear in the DMD
eigenmode.}   

\label{fig:fig5}
\end{figure}

\begin{figure}
    \centering
        \includegraphics[scale=0.170]{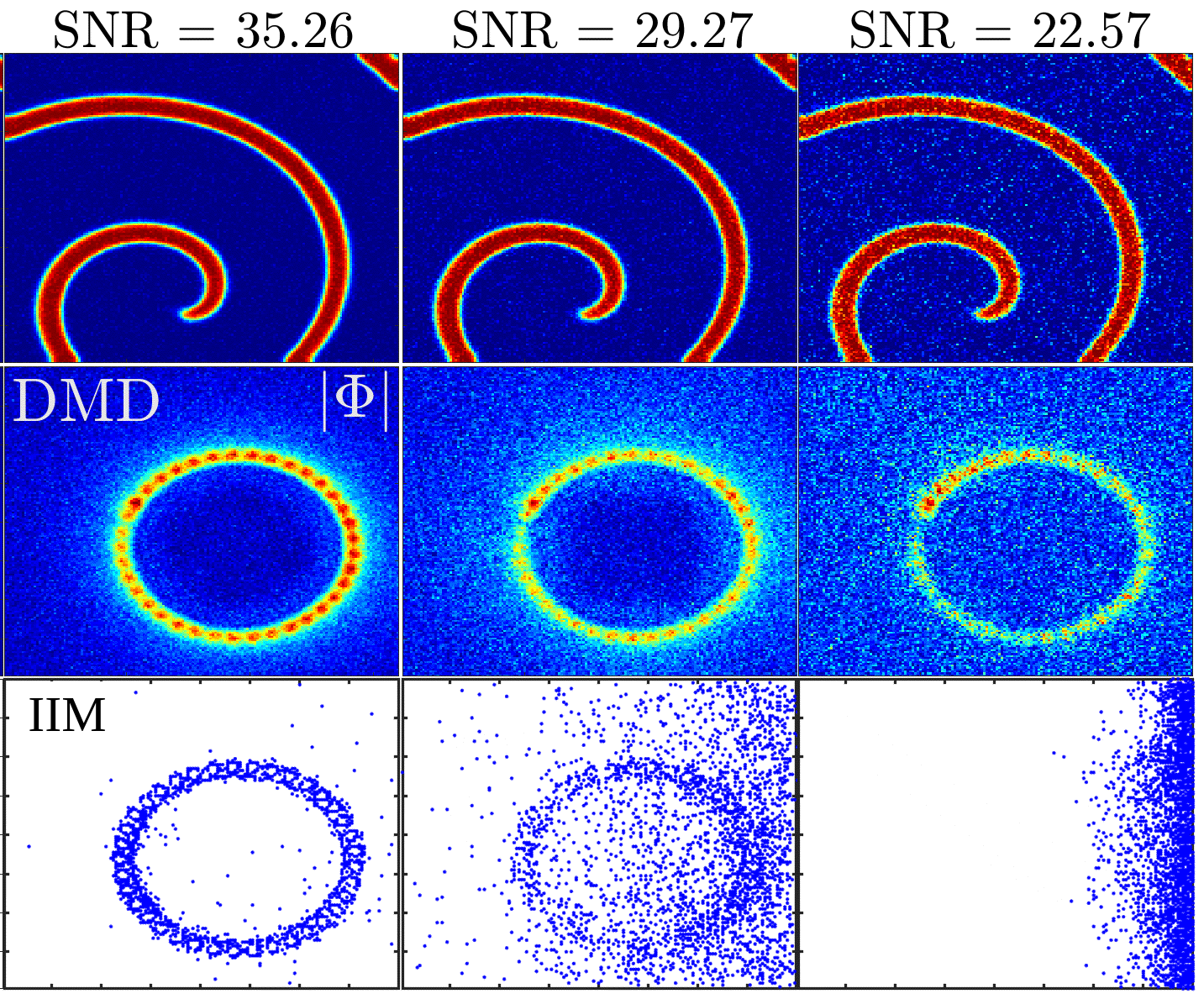}

\caption{Pseudocolor plots of (top row) $u$ for the Barkley model 
[Eqs.~(\ref{B1}) and (~\ref{B2})] and (middle row) the modulus of
a DMD eigenmode $\Phi$
that displays  high intensity along the tip trajectory; these plots
are for cases with external noise and different values of the signal-to-noise 
ratio (SNR). (Bottom row) TTs tracked by the IIM; the
TTs from the IIM are more sensitive to external noise than
their counterparts in the DMD eigenmode.}   
   
	\label{fig:fig4}
\end{figure}

\subsection{Spiral TT in the presence of heterogeneities in the medium or
noise} 
\label{subsec:heteroTT}

We use the O'Hara-Rudy model \cite{o2011simulation} of cardiac
excitation waves for our study of TTs in a heterogeneous medium. 
Figure~\ref{fig:fig5} shows how DMD and IIMs track TT in the medium with two different
percentages $P_o$ of obstacles. We find that, although both methods can locate
the region where the TT is confined, the TT plot from the IIM is
associated with randomly scattered points in the background, which are
suppressed in the TT pattern extracted by the DMD eigenmode. Moreover, we
check how these two methods perform in the presence of noise in the signal.
Such noise can arise in the data-collection processes in real experiments.
Figure~\ref{fig:fig4} shows TTs for three different values of signal-to-noise
(SNR). It shows that IIM is more sensitive to noise and it fails
to track the TT for SNR $<22$, whereas DMD can still capture the TT pattern.
DMD can produce the TT pattern upto SNR$\simeq 16$. In summary,
our results demonstrate that, with external noise,  DMDTT is a
more robust and versatile method for tracking TTs than IIM.


\begin{figure*}
   \centering
       \includegraphics[scale=0.205]{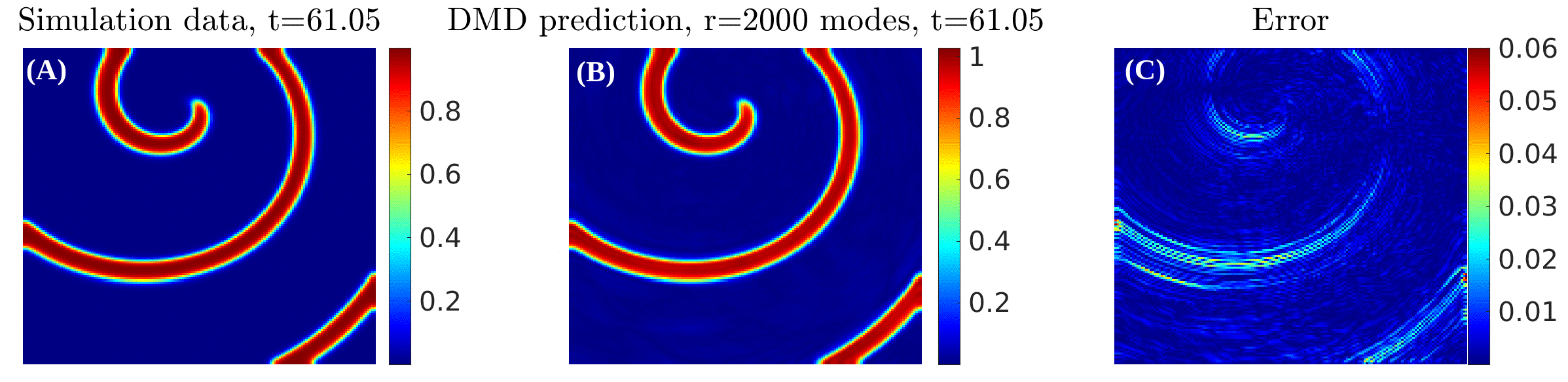}

\caption{Pseudocolor plots at a representative time of (A) $u$, for the
Barkley model [Eqs.~(\ref{B1}) and (\ref{B2})], showing a spiral wave
from our simulations; (B) the DMD-reconstructed spiral wave ($u^P$ in
Eq.~(\ref{eq:recon}); and (C) the error $u - u^P$ in the DMD-based
prediction. Movie M1 in the Supplemental Material~\cite{supmat} shows
the complete spatiotemporal evolutions of (A), (B), and (C).}

    \label{fig:fig7}
\end{figure*}

\subsection{DMD prediction of spiral-wave dynamics}
\label{subsec:reconstruct}

The DMD technique can also be used to reconstruct
(approximately), and hence predict, the spatiotemporal evolution of the spiral
waves in the models we consider. 
For this reconstruction we require the
eigenmodes $\Phi_i$, their eigenvalues $\lambda_i$, and the
amplitudes $b_i$ associated with every mode $\Phi_i$ (see, e.g.,
Ref.~\cite{tu2013dynamic}).  We illustrate this for the Barkley model
[Eqs.~(\ref{B1}) and (\ref{B2})], where we predict the spiral-wave dynamics as
follows; at any instant of time $t$, the predicted solution
$u^P{(\bf{r},t)}$ is 
\begin{equation}
	u^P({\bf{r}},t) = \sum_{i=1}^{m} b_i \Phi_i({\bf{r}}) \lambda_i^{t}, 
\label{eq:recon}
\end{equation} 
where the coefficient $b_i = (\Phi^{\dag} \Phi)^{-1}
\Phi^{\dag}({\bf{r}}) x_0$, the $\dag$ denotes Hermitian adjoint,
$m$ is the number of columns in $X_1$, and $x_0$ is
the first column in $X_1$ (Eq.(~\ref{eq:X1X2})).  We show in
Figs.~\ref{fig:fig7}(A), (B), and (C) pseudocolor plots of $u({\bf{r}},t)$ [from
Eqs.~(\ref{B1}) and (\ref{B2})], the DMD prediction $u^P({\bf{r}},t)$ [from
Eq.~(\ref{eq:recon})], and the error $u({\bf{r}})- u^P({\bf{r}},t)$,
respectively. From the plots in Figs.~\ref{fig:fig7}(A), (B), and (C) we
conclude that the DMD-based spiral-wave reconstruction works well here because
the error $u({\bf{r}},t)- u^P({\bf{r}},t) \simeq 10^{-2}$. We suggest that 
such a DMD-based prediction can be used \textit{mutatis mutandis}
with potentiometric voltage data for spiral waves from
\textit{ex-vivo} and \textit{in-vitro} experiments.

%

\section{Conclusions}
\label{sec:Conclusions}

The DMD method has been used to extract coherent structures in various
fluid-dynamical experiments~\cite{Schmid2011b,Schmid2010} and
simulations~\cite{DMDBook}. It has been applied in the study of spatially
extended systems to analyze spatiotemporal patterns emerging from the evolution
of nonlinear PDEs~\cite{DMDBook}.  The working principles of DMD have been
linked to the Koopman operator theory of dynamical
systems~\cite{rowley2009spectral,nathan2018applied,brunton2021modern}.
Furthermore, DMD provides a data-driven approach for reduced-order modeling of
high-rank dynamical
systems~\cite{Schmid2010,Schmid2011,Schmid2011b,nathan2018applied,DMDBook,krake2021visualization,tu2013dynamic}.

We have demonstrated that DMD provides a powerful method for (a) the detection
of spiral-wave TT patterns and (b) spiral-wave reconstruction in excitable
media such as cardiac tissue, for which we employ the two-variable Barkley
model [Eqs.~(\ref{B1}) and (\ref{B2})] and the biophysically realistic ORd
model [Eq.~(\ref{eq:VmPDE})]. Such a detailed application of the DMD method to
the study of spiral-wave evolution in excitable media has not been attempted
hitherto. [Two studies have applied DMD to spiral waves: one discusses the
extraction of an approximate governing equation for the spiral
waves~\cite{champion2019data}; and the other extracts observables that are
possible candidates for Koopman operators~\cite{nathan2018applied}.] Our
application of DMD to spiral waves in mathematical models for excitable media
and cardiac tissue shows leads to new insights into spiral-tip trajectories and
the prediction of the dynamics of these waves. Furthermore, our methods can be
used, in both experimental and numerical investigations, of such waves in all
excitable and oscillatory media.

We have carried out a comparison of the conventional IIM and our DMD-based TT
for mathemtatical models of cardiac tissue in (a) a homogeneous medium, (2)
with heterogeneities in the medium, and  (c) in the presence of external noise
in the signal.  We find that both DMDTT and IIM can track various patterns,
including the circular and complicated ones shown in Figs.~\ref{fig:fig1} and
~\ref{fig:fig2}, if the sampling interval $\tau$ is small.  However, for a
large value of $\tau$, the IIM fails to track the TT, whereas DMD can still
capture the TT pattern. We show, furthermore (Fig.~\ref{fig:fig3}), that DMD
can be used to locate (a) phase singularities, even when there are multiple
spiral waves, and (b) the domain boundaries between different spiral waves. In
a medium with heterogeneities, both DMD and IIM can track TTs; however, TT
plots from the IIM can show randomly scattered points in the background, which
are suppressed in the TT patterns we obtain via DMD. Finally, in the presence
of external noise in the signal, which can be present in experimental data, we
show that IIM fails to track TT for the signal-to-noise ratio SNR $< 22$; by
contrast, our DMD method can capture TT patterns up until SNR $\simeq 16$, so
DMD provides a more robust method to track TTs, in the presence of noise, as
compared to IIM.

We have noted already that the accurate tracking of the tip of
a spiral wave and the mapping of phase singularities can give valuable
information about its evolution of the spiral
waves~\cite{aronis2017rotors,umapathy2010phase,bray2001experimental,gray1998spatial}.
Complicated  meandering TT patterns are vulnerable to spiral-wave
instabilities~\cite{qu2000origins,fenton2002multiple};
TTs can provide insights into the underlying mechanisms of transitions 
from single- to multiple-spiral states, which are of great interest 
in the study of cardiac arryhthmias~\cite{fenton2002multiple}. Therefore, 
it is important to develop versatile methods for tracking TTs;
these methods should be applicable in varied experimental settings. 
We have shown, by using three settings, that DMDTT 
can be more versatile than the commonly used IIM for tracking
TTs. 

We expect that the DMD methods, which we have elucidated above, can be used to
study electrical-activation patterns in mammalian hearts, at least in
\textit{ex-vivo} optical-mapping experiments with Lagendroff-perfused hearts.
These methods can be applied on a set of optical images, collected successively
at certain intervals of time, for the detection of phase singularities.  The
precise location of such phase singularities can be used for accurate ablation,
which can help in terminating life-threatening cardiac arrhythmias.
Furthermore, our DMD methods can be used fruitfully for such singularity
detection in conjunction with conventional phase-singularity-mapping
methods~\cite{bray2001experimental,laughner2012processing}.

We have demonstrated how to carry out spiral-wave reconstruction in excitable
media, such as cardiac tissue, by using the DMD eigenmodes $\Phi_i$  [see
Eq.(~\ref{eq:recon}) and Fig.~\ref{fig:fig7}]. Such a DMD-based prediction can
be used \textit{mutatis mutandis} with experimental data for spiral waves of
electrical activation from \textit{ex-vivo} and \textit{in-vitro}
experiments. We hope our work will lead to such experimental investigations,
which have the potential to play an important role in the field of
life-threatening cardiac arrhythmias.




We end our paper by discussing some limitations of our study. Here, we have
focused only on the detection of the phase singularity of a spiral wave and its
TT; however, other forms of activation patterns have been implicated in the
occurrence of arrhythmias, such as focal or multiple-wave activation
patterns~\cite{miragoli2007myofibroblasts,zimik2015computational,alonso2016reentry}.
In future work we will conduct a detailed analysis of how DMD can be used to
characterize such activation patterns and how DMD can discriminate such
patterns from spiral waves. Moreover, our study is restricted to
two-dimensions. We  will extended this to three-dimensional and anatomically
realistic domains, which display rich forms of spatiotemporal organizations
like scroll waves of
excitation (see, e.g., Refs.~\cite{maucher2017length,maucher2019dynamics,weingard2017scroll,rajany2021effects}).

\begin{acknowledgments}

We thank SERC (IISc) for computational resources and CSIR, SERB, and
the National Supercomputing Mission (NSM) for the support.

\end{acknowledgments}

\bibliography{references}

\providecommand{\noopsort}[1]{}\providecommand{\singleletter}[1]{#1}%
\begin{thebibliography}{45}%
\makeatletter
\providecommand \@ifxundefined [1]{%
 \@ifx{#1\undefined}
}%
\providecommand \@ifnum [1]{%
 \ifnum #1\expandafter \@firstoftwo
 \else \expandafter \@secondoftwo
 \fi
}%
\providecommand \@ifx [1]{%
 \ifx #1\expandafter \@firstoftwo
 \else \expandafter \@secondoftwo
 \fi
}%
\providecommand \natexlab [1]{#1}%
\providecommand \enquote  [1]{``#1''}%
\providecommand \bibnamefont  [1]{#1}%
\providecommand \bibfnamefont [1]{#1}%
\providecommand \citenamefont [1]{#1}%
\providecommand \href@noop [0]{\@secondoftwo}%
\providecommand \href [0]{\begingroup \@sanitize@url \@href}%
\providecommand \@href[1]{\@@startlink{#1}\@@href}%
\providecommand \@@href[1]{\endgroup#1\@@endlink}%
\providecommand \@sanitize@url [0]{\catcode `\\12\catcode `\$12\catcode
  `\&12\catcode `\#12\catcode `\^12\catcode `\_12\catcode `\%12\relax}%
\providecommand \@@startlink[1]{}%
\providecommand \@@endlink[0]{}%
\providecommand \url  [0]{\begingroup\@sanitize@url \@url }%
\providecommand \@url [1]{\endgroup\@href {#1}{\urlprefix }}%
\providecommand \urlprefix  [0]{URL }%
\providecommand \Eprint [0]{\href }%
\providecommand \doibase [0]{https://doi.org/}%
\providecommand \selectlanguage [0]{\@gobble}%
\providecommand \bibinfo  [0]{\@secondoftwo}%
\providecommand \bibfield  [0]{\@secondoftwo}%
\providecommand \translation [1]{[#1]}%
\providecommand \BibitemOpen [0]{}%
\providecommand \bibitemStop [0]{}%
\providecommand \bibitemNoStop [0]{.\EOS\space}%
\providecommand \EOS [0]{\spacefactor3000\relax}%
\providecommand \BibitemShut  [1]{\csname bibitem#1\endcsname}%
\let\auto@bib@innerbib\@empty
\bibitem [{\citenamefont {Winfree}(1972)}]{winfree1972spiral}%
  \BibitemOpen
  \bibfield  {author} {\bibinfo {author} {\bibfnamefont {A.~T.}\ \bibnamefont
  {Winfree}},\ }\bibfield  {title} {\bibinfo {title} {Spiral waves of chemical
  activity},\ }\href@noop {} {\bibfield  {journal} {\bibinfo  {journal}
  {Science}\ }\textbf {\bibinfo {volume} {175}},\ \bibinfo {pages} {634}
  (\bibinfo {year} {1972})}\BibitemShut {NoStop}%
\bibitem [{\citenamefont {Lechleiter}\ \emph {et~al.}(1991)\citenamefont
  {Lechleiter}, \citenamefont {Girard}, \citenamefont {Peralta},\ and\
  \citenamefont {Clapham}}]{lechleiter1991spiral}%
  \BibitemOpen
  \bibfield  {author} {\bibinfo {author} {\bibfnamefont {J.}~\bibnamefont
  {Lechleiter}}, \bibinfo {author} {\bibfnamefont {S.}~\bibnamefont {Girard}},
  \bibinfo {author} {\bibfnamefont {E.}~\bibnamefont {Peralta}},\ and\ \bibinfo
  {author} {\bibfnamefont {D.}~\bibnamefont {Clapham}},\ }\bibfield  {title}
  {\bibinfo {title} {Spiral calcium wave propagation and annihilation in
  xenopus laevis oocytes},\ }\href@noop {} {\bibfield  {journal} {\bibinfo
  {journal} {Science}\ }\textbf {\bibinfo {volume} {252}},\ \bibinfo {pages}
  {123} (\bibinfo {year} {1991})}\BibitemShut {NoStop}%
\bibitem [{\citenamefont {Tyson}\ and\ \citenamefont
  {Murray}(1989)}]{tyson1989cyclic}%
  \BibitemOpen
  \bibfield  {author} {\bibinfo {author} {\bibfnamefont {J.~J.}\ \bibnamefont
  {Tyson}}\ and\ \bibinfo {author} {\bibfnamefont {J.}~\bibnamefont {Murray}},\
  }\bibfield  {title} {\bibinfo {title} {Cyclic amp waves during aggregation of
  dictyostelium amoebae},\ }\href@noop {} {\bibfield  {journal} {\bibinfo
  {journal} {Development}\ }\textbf {\bibinfo {volume} {106}},\ \bibinfo
  {pages} {421} (\bibinfo {year} {1989})}\BibitemShut {NoStop}%
\bibitem [{\citenamefont {Falcke}\ \emph {et~al.}(1992)\citenamefont {Falcke},
  \citenamefont {B{\"a}r}, \citenamefont {Engel},\ and\ \citenamefont
  {Eiswirth}}]{falcke1992traveling}%
  \BibitemOpen
  \bibfield  {author} {\bibinfo {author} {\bibfnamefont {M.}~\bibnamefont
  {Falcke}}, \bibinfo {author} {\bibfnamefont {M.}~\bibnamefont {B{\"a}r}},
  \bibinfo {author} {\bibfnamefont {H.}~\bibnamefont {Engel}},\ and\ \bibinfo
  {author} {\bibfnamefont {M.}~\bibnamefont {Eiswirth}},\ }\bibfield  {title}
  {\bibinfo {title} {Traveling waves in the co oxidation on pt (110): Theory},\
  }\href@noop {} {\bibfield  {journal} {\bibinfo  {journal} {The Journal of
  chemical physics}\ }\textbf {\bibinfo {volume} {97}},\ \bibinfo {pages}
  {4555} (\bibinfo {year} {1992})}\BibitemShut {NoStop}%
\bibitem [{\citenamefont {Seiden}\ and\ \citenamefont
  {Curland}(2015)}]{seiden2015tongue}%
  \BibitemOpen
  \bibfield  {author} {\bibinfo {author} {\bibfnamefont {G.}~\bibnamefont
  {Seiden}}\ and\ \bibinfo {author} {\bibfnamefont {S.}~\bibnamefont
  {Curland}},\ }\bibfield  {title} {\bibinfo {title} {The tongue as an
  excitable medium},\ }\href@noop {} {\bibfield  {journal} {\bibinfo  {journal}
  {New Journal of Physics}\ }\textbf {\bibinfo {volume} {17}},\ \bibinfo
  {pages} {033049} (\bibinfo {year} {2015})}\BibitemShut {NoStop}%
\bibitem [{\citenamefont {McGuire}\ \emph {et~al.}(2021)\citenamefont
  {McGuire}, \citenamefont {Fuller}, \citenamefont {Lindner},\ and\
  \citenamefont {Manz}}]{mcguire2021geographic}%
  \BibitemOpen
  \bibfield  {author} {\bibinfo {author} {\bibfnamefont {M.~K.}\ \bibnamefont
  {McGuire}}, \bibinfo {author} {\bibfnamefont {C.~A.}\ \bibnamefont {Fuller}},
  \bibinfo {author} {\bibfnamefont {J.~F.}\ \bibnamefont {Lindner}},\ and\
  \bibinfo {author} {\bibfnamefont {N.}~\bibnamefont {Manz}},\ }\bibfield
  {title} {\bibinfo {title} {Geographic tongue as a reaction--diffusion
  system},\ }\href@noop {} {\bibfield  {journal} {\bibinfo  {journal} {Chaos:
  An Interdisciplinary Journal of Nonlinear Science}\ }\textbf {\bibinfo
  {volume} {31}},\ \bibinfo {pages} {033118} (\bibinfo {year}
  {2021})}\BibitemShut {NoStop}%
\bibitem [{\citenamefont {Marino}\ and\ \citenamefont
  {Giacomelli}(2019)}]{marino2019excitable}%
  \BibitemOpen
  \bibfield  {author} {\bibinfo {author} {\bibfnamefont {F.}~\bibnamefont
  {Marino}}\ and\ \bibinfo {author} {\bibfnamefont {G.}~\bibnamefont
  {Giacomelli}},\ }\bibfield  {title} {\bibinfo {title} {Excitable wave
  patterns in temporal systems with two long delays and their observation in a
  semiconductor laser experiment},\ }\href@noop {} {\bibfield  {journal}
  {\bibinfo  {journal} {Physical Review Letters}\ }\textbf {\bibinfo {volume}
  {122}},\ \bibinfo {pages} {174102} (\bibinfo {year} {2019})}\BibitemShut
  {NoStop}%
\bibitem [{\citenamefont {Allessie}\ \emph {et~al.}(1977)\citenamefont
  {Allessie}, \citenamefont {Bonke},\ and\ \citenamefont
  {Schopman}}]{allessie1977circus}%
  \BibitemOpen
  \bibfield  {author} {\bibinfo {author} {\bibfnamefont {M.~A.}\ \bibnamefont
  {Allessie}}, \bibinfo {author} {\bibfnamefont {F.}~\bibnamefont {Bonke}},\
  and\ \bibinfo {author} {\bibfnamefont {F.}~\bibnamefont {Schopman}},\
  }\bibfield  {title} {\bibinfo {title} {Circus movement in rabbit atrial
  muscle as a mechanism of tachycardia. iii. the" leading circle" concept: a
  new model of circus movement in cardiac tissue without the involvement of an
  anatomical obstacle.},\ }\href@noop {} {\bibfield  {journal} {\bibinfo
  {journal} {Circulation research}\ }\textbf {\bibinfo {volume} {41}},\
  \bibinfo {pages} {9} (\bibinfo {year} {1977})}\BibitemShut {NoStop}%
\bibitem [{\citenamefont {Davidenko}\ \emph {et~al.}(1990)\citenamefont
  {Davidenko}, \citenamefont {Kent}, \citenamefont {Chialvo}, \citenamefont
  {Michaels},\ and\ \citenamefont {Jalife}}]{davidenko1990sustained}%
  \BibitemOpen
  \bibfield  {author} {\bibinfo {author} {\bibfnamefont {J.~M.}\ \bibnamefont
  {Davidenko}}, \bibinfo {author} {\bibfnamefont {P.~F.}\ \bibnamefont {Kent}},
  \bibinfo {author} {\bibfnamefont {D.~R.}\ \bibnamefont {Chialvo}}, \bibinfo
  {author} {\bibfnamefont {D.~C.}\ \bibnamefont {Michaels}},\ and\ \bibinfo
  {author} {\bibfnamefont {J.}~\bibnamefont {Jalife}},\ }\bibfield  {title}
  {\bibinfo {title} {Sustained vortex-like waves in normal isolated ventricular
  muscle.},\ }\href@noop {} {\bibfield  {journal} {\bibinfo  {journal}
  {Proceedings of the National Academy of Sciences}\ }\textbf {\bibinfo
  {volume} {87}},\ \bibinfo {pages} {8785} (\bibinfo {year}
  {1990})}\BibitemShut {NoStop}%
\bibitem [{\citenamefont {Pertsov}\ \emph {et~al.}(1993)\citenamefont
  {Pertsov}, \citenamefont {Davidenko}, \citenamefont {Salomonsz},
  \citenamefont {Baxter},\ and\ \citenamefont {Jalife}}]{pertsov1993spiral}%
  \BibitemOpen
  \bibfield  {author} {\bibinfo {author} {\bibfnamefont {A.~M.}\ \bibnamefont
  {Pertsov}}, \bibinfo {author} {\bibfnamefont {J.~M.}\ \bibnamefont
  {Davidenko}}, \bibinfo {author} {\bibfnamefont {R.}~\bibnamefont
  {Salomonsz}}, \bibinfo {author} {\bibfnamefont {W.~T.}\ \bibnamefont
  {Baxter}},\ and\ \bibinfo {author} {\bibfnamefont {J.}~\bibnamefont
  {Jalife}},\ }\bibfield  {title} {\bibinfo {title} {Spiral waves of excitation
  underlie reentrant activity in isolated cardiac muscle.},\ }\href@noop {}
  {\bibfield  {journal} {\bibinfo  {journal} {Circulation research}\ }\textbf
  {\bibinfo {volume} {72}},\ \bibinfo {pages} {631} (\bibinfo {year}
  {1993})}\BibitemShut {NoStop}%
\bibitem [{\citenamefont {Gray}\ \emph {et~al.}(1995)\citenamefont {Gray},
  \citenamefont {Jalife}, \citenamefont {Panfilov}, \citenamefont {Baxter},
  \citenamefont {Cabo}, \citenamefont {Davidenko}, \citenamefont {Pertsov},
  \citenamefont {Hogeweg},\ and\ \citenamefont {Winfree}}]{gray1995mechanisms}%
  \BibitemOpen
  \bibfield  {author} {\bibinfo {author} {\bibfnamefont {R.~A.}\ \bibnamefont
  {Gray}}, \bibinfo {author} {\bibfnamefont {J.}~\bibnamefont {Jalife}},
  \bibinfo {author} {\bibfnamefont {A.~V.}\ \bibnamefont {Panfilov}}, \bibinfo
  {author} {\bibfnamefont {W.~T.}\ \bibnamefont {Baxter}}, \bibinfo {author}
  {\bibfnamefont {C.}~\bibnamefont {Cabo}}, \bibinfo {author} {\bibfnamefont
  {J.~M.}\ \bibnamefont {Davidenko}}, \bibinfo {author} {\bibfnamefont {A.~M.}\
  \bibnamefont {Pertsov}}, \bibinfo {author} {\bibfnamefont {P.}~\bibnamefont
  {Hogeweg}},\ and\ \bibinfo {author} {\bibfnamefont {A.~T.}\ \bibnamefont
  {Winfree}},\ }\bibfield  {title} {\bibinfo {title} {Mechanisms of cardiac
  fibrillation},\ }\href@noop {} {\bibfield  {journal} {\bibinfo  {journal}
  {Science}\ }\textbf {\bibinfo {volume} {270}},\ \bibinfo {pages} {1222}
  (\bibinfo {year} {1995})}\BibitemShut {NoStop}%
\bibitem [{\citenamefont {Gray}\ \emph {et~al.}(1998)\citenamefont {Gray},
  \citenamefont {Pertsov},\ and\ \citenamefont {Jalife}}]{gray1998spatial}%
  \BibitemOpen
  \bibfield  {author} {\bibinfo {author} {\bibfnamefont {R.~A.}\ \bibnamefont
  {Gray}}, \bibinfo {author} {\bibfnamefont {A.~M.}\ \bibnamefont {Pertsov}},\
  and\ \bibinfo {author} {\bibfnamefont {J.}~\bibnamefont {Jalife}},\
  }\bibfield  {title} {\bibinfo {title} {Spatial and temporal organization
  during cardiac fibrillation},\ }\href@noop {} {\bibfield  {journal} {\bibinfo
   {journal} {Nature}\ }\textbf {\bibinfo {volume} {392}},\ \bibinfo {pages}
  {75} (\bibinfo {year} {1998})}\BibitemShut {NoStop}%
\bibitem [{\citenamefont {Fenton}\ and\ \citenamefont
  {Karma}(1998)}]{fenton1998vortex}%
  \BibitemOpen
  \bibfield  {author} {\bibinfo {author} {\bibfnamefont {F.}~\bibnamefont
  {Fenton}}\ and\ \bibinfo {author} {\bibfnamefont {A.}~\bibnamefont {Karma}},\
  }\bibfield  {title} {\bibinfo {title} {Vortex dynamics in three-dimensional
  continuous myocardium with fiber rotation: Filament instability and
  fibrillation},\ }\href@noop {} {\bibfield  {journal} {\bibinfo  {journal}
  {Chaos: An Interdisciplinary Journal of Nonlinear Science}\ }\textbf
  {\bibinfo {volume} {8}},\ \bibinfo {pages} {20} (\bibinfo {year}
  {1998})}\BibitemShut {NoStop}%
\bibitem [{\citenamefont {Qu}\ \emph {et~al.}(2000)\citenamefont {Qu},
  \citenamefont {Xie}, \citenamefont {Garfinkel},\ and\ \citenamefont
  {Weiss}}]{qu2000origins}%
  \BibitemOpen
  \bibfield  {author} {\bibinfo {author} {\bibfnamefont {Z.}~\bibnamefont
  {Qu}}, \bibinfo {author} {\bibfnamefont {F.}~\bibnamefont {Xie}}, \bibinfo
  {author} {\bibfnamefont {A.}~\bibnamefont {Garfinkel}},\ and\ \bibinfo
  {author} {\bibfnamefont {J.~N.}\ \bibnamefont {Weiss}},\ }\bibfield  {title}
  {\bibinfo {title} {Origins of spiral wave meander and breakup in a
  two-dimensional cardiac tissue model},\ }\href@noop {} {\bibfield  {journal}
  {\bibinfo  {journal} {Annals of biomedical engineering}\ }\textbf {\bibinfo
  {volume} {28}},\ \bibinfo {pages} {755} (\bibinfo {year} {2000})}\BibitemShut
  {NoStop}%
\bibitem [{\citenamefont {Gray}\ \emph {et~al.}(2009)\citenamefont {Gray},
  \citenamefont {Wikswo},\ and\ \citenamefont {Otani}}]{gray2009origin}%
  \BibitemOpen
  \bibfield  {author} {\bibinfo {author} {\bibfnamefont {R.~A.}\ \bibnamefont
  {Gray}}, \bibinfo {author} {\bibfnamefont {J.~P.}\ \bibnamefont {Wikswo}},\
  and\ \bibinfo {author} {\bibfnamefont {N.~F.}\ \bibnamefont {Otani}},\
  }\bibfield  {title} {\bibinfo {title} {Origin choice and petal loss in the
  flower garden of spiral wave tip trajectories},\ }\href@noop {} {\bibfield
  {journal} {\bibinfo  {journal} {Chaos: An Interdisciplinary Journal of
  Nonlinear Science}\ }\textbf {\bibinfo {volume} {19}},\ \bibinfo {pages}
  {033118} (\bibinfo {year} {2009})}\BibitemShut {NoStop}%
\bibitem [{\citenamefont {Aronis}\ \emph {et~al.}(2017)\citenamefont {Aronis},
  \citenamefont {Berger},\ and\ \citenamefont {Ashikaga}}]{aronis2017rotors}%
  \BibitemOpen
  \bibfield  {author} {\bibinfo {author} {\bibfnamefont {K.~N.}\ \bibnamefont
  {Aronis}}, \bibinfo {author} {\bibfnamefont {R.~D.}\ \bibnamefont {Berger}},\
  and\ \bibinfo {author} {\bibfnamefont {H.}~\bibnamefont {Ashikaga}},\
  }\href@noop {} {\bibinfo {title} {Rotors: how do we know when they are
  real?}} (\bibinfo {year} {2017})\BibitemShut {NoStop}%
\bibitem [{\citenamefont {Umapathy}\ \emph {et~al.}(2010)\citenamefont
  {Umapathy}, \citenamefont {Nair}, \citenamefont {Masse}, \citenamefont
  {Krishnan}, \citenamefont {Rogers}, \citenamefont {Nash},\ and\ \citenamefont
  {Nanthakumar}}]{umapathy2010phase}%
  \BibitemOpen
  \bibfield  {author} {\bibinfo {author} {\bibfnamefont {K.}~\bibnamefont
  {Umapathy}}, \bibinfo {author} {\bibfnamefont {K.}~\bibnamefont {Nair}},
  \bibinfo {author} {\bibfnamefont {S.}~\bibnamefont {Masse}}, \bibinfo
  {author} {\bibfnamefont {S.}~\bibnamefont {Krishnan}}, \bibinfo {author}
  {\bibfnamefont {J.}~\bibnamefont {Rogers}}, \bibinfo {author} {\bibfnamefont
  {M.~P.}\ \bibnamefont {Nash}},\ and\ \bibinfo {author} {\bibfnamefont
  {K.}~\bibnamefont {Nanthakumar}},\ }\bibfield  {title} {\bibinfo {title}
  {Phase mapping of cardiac fibrillation},\ }\href@noop {} {\bibfield
  {journal} {\bibinfo  {journal} {Circulation: Arrhythmia and
  Electrophysiology}\ }\textbf {\bibinfo {volume} {3}},\ \bibinfo {pages} {105}
  (\bibinfo {year} {2010})}\BibitemShut {NoStop}%
\bibitem [{\citenamefont {BRAY}\ \emph {et~al.}(2001)\citenamefont {BRAY},
  \citenamefont {LIN}, \citenamefont {Aliev}, \citenamefont {Roth},\ and\
  \citenamefont {Wikswo~Jr}}]{bray2001experimental}%
  \BibitemOpen
  \bibfield  {author} {\bibinfo {author} {\bibfnamefont {M.-A.}\ \bibnamefont
  {BRAY}}, \bibinfo {author} {\bibfnamefont {S.-F.}\ \bibnamefont {LIN}},
  \bibinfo {author} {\bibfnamefont {R.~R.}\ \bibnamefont {Aliev}}, \bibinfo
  {author} {\bibfnamefont {B.~J.}\ \bibnamefont {Roth}},\ and\ \bibinfo
  {author} {\bibfnamefont {J.~P.}\ \bibnamefont {Wikswo~Jr}},\ }\bibfield
  {title} {\bibinfo {title} {Experimental and theoretical analysis of phase
  singularity dynamics in cardiac tissue},\ }\href@noop {} {\bibfield
  {journal} {\bibinfo  {journal} {Journal of cardiovascular electrophysiology}\
  }\textbf {\bibinfo {volume} {12}},\ \bibinfo {pages} {716} (\bibinfo {year}
  {2001})}\BibitemShut {NoStop}%
\bibitem [{\citenamefont {Schmid}(2010)}]{Schmid2010}%
  \BibitemOpen
  \bibfield  {author} {\bibinfo {author} {\bibfnamefont {P.~J.}\ \bibnamefont
  {Schmid}},\ }\bibfield  {title} {\bibinfo {title} {Dynamic mode decomposition
  of numerical and experimental data},\ }\href@noop {} {\bibfield  {journal}
  {\bibinfo  {journal} {Journal of fluid mechanics}\ }\textbf {\bibinfo
  {volume} {656}},\ \bibinfo {pages} {5} (\bibinfo {year} {2010})}\BibitemShut
  {NoStop}%
\bibitem [{\citenamefont {Schmid}\ \emph {et~al.}(2011)\citenamefont {Schmid},
  \citenamefont {Li}, \citenamefont {Juniper},\ and\ \citenamefont
  {Pust}}]{Schmid2011}%
  \BibitemOpen
  \bibfield  {author} {\bibinfo {author} {\bibfnamefont {P.~J.}\ \bibnamefont
  {Schmid}}, \bibinfo {author} {\bibfnamefont {L.}~\bibnamefont {Li}}, \bibinfo
  {author} {\bibfnamefont {M.~P.}\ \bibnamefont {Juniper}},\ and\ \bibinfo
  {author} {\bibfnamefont {O.}~\bibnamefont {Pust}},\ }\bibfield  {title}
  {\bibinfo {title} {Applications of the dynamic mode decomposition},\
  }\href@noop {} {\bibfield  {journal} {\bibinfo  {journal} {Theoretical and
  Computational Fluid Dynamics}\ }\textbf {\bibinfo {volume} {25}},\ \bibinfo
  {pages} {249} (\bibinfo {year} {2011})}\BibitemShut {NoStop}%
\bibitem [{\citenamefont {Schmid}(2011)}]{Schmid2011b}%
  \BibitemOpen
  \bibfield  {author} {\bibinfo {author} {\bibfnamefont {P.~J.}\ \bibnamefont
  {Schmid}},\ }\bibfield  {title} {\bibinfo {title} {Application of the dynamic
  mode decomposition to experimental data},\ }\href@noop {} {\bibfield
  {journal} {\bibinfo  {journal} {Experiments in fluids}\ }\textbf {\bibinfo
  {volume} {50}},\ \bibinfo {pages} {1123} (\bibinfo {year}
  {2011})}\BibitemShut {NoStop}%
\bibitem [{\citenamefont {Nathan~Kutz}\ \emph {et~al.}(2018)\citenamefont
  {Nathan~Kutz}, \citenamefont {Proctor},\ and\ \citenamefont
  {Brunton}}]{nathan2018applied}%
  \BibitemOpen
  \bibfield  {author} {\bibinfo {author} {\bibfnamefont {J.}~\bibnamefont
  {Nathan~Kutz}}, \bibinfo {author} {\bibfnamefont {J.~L.}\ \bibnamefont
  {Proctor}},\ and\ \bibinfo {author} {\bibfnamefont {S.~L.}\ \bibnamefont
  {Brunton}},\ }\bibfield  {title} {\bibinfo {title} {Applied koopman theory
  for partial differential equations and data-driven modeling of
  spatio-temporal systems},\ }\href@noop {} {\bibfield  {journal} {\bibinfo
  {journal} {Complexity}\ }\textbf {\bibinfo {volume} {2018}} (\bibinfo {year}
  {2018})}\BibitemShut {NoStop}%
\bibitem [{\citenamefont {Kutz}\ \emph {et~al.}(2016)\citenamefont {Kutz},
  \citenamefont {Brunton}, \citenamefont {Brunton},\ and\ \citenamefont
  {Proctor}}]{DMDBook}%
  \BibitemOpen
  \bibfield  {author} {\bibinfo {author} {\bibfnamefont {J.~N.}\ \bibnamefont
  {Kutz}}, \bibinfo {author} {\bibfnamefont {S.~L.}\ \bibnamefont {Brunton}},
  \bibinfo {author} {\bibfnamefont {B.~W.}\ \bibnamefont {Brunton}},\ and\
  \bibinfo {author} {\bibfnamefont {J.~L.}\ \bibnamefont {Proctor}},\
  }\href@noop {} {\emph {\bibinfo {title} {Dynamic mode decomposition:
  data-driven modeling of complex systems}}}\ (\bibinfo  {publisher} {SIAM},\
  \bibinfo {year} {2016})\BibitemShut {NoStop}%
\bibitem [{\citenamefont {Krake}\ \emph {et~al.}(2021)\citenamefont {Krake},
  \citenamefont {Reinhardt}, \citenamefont {Hlawatsch}, \citenamefont
  {Eberhardt},\ and\ \citenamefont {Weiskopf}}]{krake2021visualization}%
  \BibitemOpen
  \bibfield  {author} {\bibinfo {author} {\bibfnamefont {T.}~\bibnamefont
  {Krake}}, \bibinfo {author} {\bibfnamefont {S.}~\bibnamefont {Reinhardt}},
  \bibinfo {author} {\bibfnamefont {M.}~\bibnamefont {Hlawatsch}}, \bibinfo
  {author} {\bibfnamefont {B.}~\bibnamefont {Eberhardt}},\ and\ \bibinfo
  {author} {\bibfnamefont {D.}~\bibnamefont {Weiskopf}},\ }\bibfield  {title}
  {\bibinfo {title} {Visualization and selection of dynamic mode decomposition
  components for unsteady flow},\ }\href@noop {} {\bibfield  {journal}
  {\bibinfo  {journal} {Visual Informatics}\ }\textbf {\bibinfo {volume} {5}},\
  \bibinfo {pages} {15} (\bibinfo {year} {2021})}\BibitemShut {NoStop}%
\bibitem [{\citenamefont {Tu}(2013)}]{tu2013dynamic}%
  \BibitemOpen
  \bibfield  {author} {\bibinfo {author} {\bibfnamefont {J.~H.}\ \bibnamefont
  {Tu}},\ }\emph {\bibinfo {title} {Dynamic mode decomposition: Theory and
  applications}},\ \href@noop {} {Ph.D. thesis},\ \bibinfo  {school} {Princeton
  University} (\bibinfo {year} {2013})\BibitemShut {NoStop}%
\bibitem [{\citenamefont {Champion}\ \emph {et~al.}(2019)\citenamefont
  {Champion}, \citenamefont {Lusch}, \citenamefont {Kutz},\ and\ \citenamefont
  {Brunton}}]{champion2019data}%
  \BibitemOpen
  \bibfield  {author} {\bibinfo {author} {\bibfnamefont {K.}~\bibnamefont
  {Champion}}, \bibinfo {author} {\bibfnamefont {B.}~\bibnamefont {Lusch}},
  \bibinfo {author} {\bibfnamefont {J.~N.}\ \bibnamefont {Kutz}},\ and\
  \bibinfo {author} {\bibfnamefont {S.~L.}\ \bibnamefont {Brunton}},\
  }\bibfield  {title} {\bibinfo {title} {Data-driven discovery of coordinates
  and governing equations},\ }\href@noop {} {\bibfield  {journal} {\bibinfo
  {journal} {Proceedings of the National Academy of Sciences}\ }\textbf
  {\bibinfo {volume} {116}},\ \bibinfo {pages} {22445} (\bibinfo {year}
  {2019})}\BibitemShut {NoStop}%
\bibitem [{\citenamefont {Barkley}(1991)}]{Barkley1991}%
  \BibitemOpen
  \bibfield  {author} {\bibinfo {author} {\bibfnamefont {D.}~\bibnamefont
  {Barkley}},\ }\href@noop {} {\bibfield  {journal} {\bibinfo  {journal}
  {Physica D: Nonlinear Phenomena}\ }\textbf {\bibinfo {volume} {49}},\
  \bibinfo {pages} {61} (\bibinfo {year} {1991})}\BibitemShut {NoStop}%
\bibitem [{\citenamefont {O'Hara}\ \emph {et~al.}(2011)\citenamefont {O'Hara},
  \citenamefont {Vir{\'a}g}, \citenamefont {Varr{\'o}},\ and\ \citenamefont
  {Rudy}}]{o2011simulation}%
  \BibitemOpen
  \bibfield  {author} {\bibinfo {author} {\bibfnamefont {T.}~\bibnamefont
  {O'Hara}}, \bibinfo {author} {\bibfnamefont {L.}~\bibnamefont {Vir{\'a}g}},
  \bibinfo {author} {\bibfnamefont {A.}~\bibnamefont {Varr{\'o}}},\ and\
  \bibinfo {author} {\bibfnamefont {Y.}~\bibnamefont {Rudy}},\ }\bibfield
  {title} {\bibinfo {title} {Simulation of the undiseased human cardiac
  ventricular action potential: model formulation and experimental
  validation},\ }\href@noop {} {\bibfield  {journal} {\bibinfo  {journal} {PLoS
  computational biology}\ }\textbf {\bibinfo {volume} {7}},\ \bibinfo {pages}
  {e1002061} (\bibinfo {year} {2011})}\BibitemShut {NoStop}%
\bibitem [{\citenamefont {Zimik}\ \emph {et~al.}(2015)\citenamefont {Zimik},
  \citenamefont {Nayak},\ and\ \citenamefont
  {Pandit}}]{zimik2015computational}%
  \BibitemOpen
  \bibfield  {author} {\bibinfo {author} {\bibfnamefont {S.}~\bibnamefont
  {Zimik}}, \bibinfo {author} {\bibfnamefont {A.~R.}\ \bibnamefont {Nayak}},\
  and\ \bibinfo {author} {\bibfnamefont {R.}~\bibnamefont {Pandit}},\
  }\bibfield  {title} {\bibinfo {title} {A computational study of the factors
  influencing the pvc-triggering ability of a cluster of early
  afterdepolarization-capable myocytes},\ }\href@noop {} {\bibfield  {journal}
  {\bibinfo  {journal} {PloS one}\ }\textbf {\bibinfo {volume} {10}},\ \bibinfo
  {pages} {e0144979} (\bibinfo {year} {2015})}\BibitemShut {NoStop}%
\bibitem [{\citenamefont {Zimik}\ and\ \citenamefont
  {Pandit}(2017)}]{zimik2017reentry}%
  \BibitemOpen
  \bibfield  {author} {\bibinfo {author} {\bibfnamefont {S.}~\bibnamefont
  {Zimik}}\ and\ \bibinfo {author} {\bibfnamefont {R.}~\bibnamefont {Pandit}},\
  }\bibfield  {title} {\bibinfo {title} {Reentry via high-frequency pacing in a
  mathematical model for human-ventricular cardiac tissue with a localized
  fibrotic region},\ }\href@noop {} {\bibfield  {journal} {\bibinfo  {journal}
  {Scientific reports}\ }\textbf {\bibinfo {volume} {7}},\ \bibinfo {pages} {1}
  (\bibinfo {year} {2017})}\BibitemShut {NoStop}%
\bibitem [{\citenamefont {Ten~Tusscher}\ and\ \citenamefont
  {Panfilov}(2007)}]{ten2007influence}%
  \BibitemOpen
  \bibfield  {author} {\bibinfo {author} {\bibfnamefont {K.~H.}\ \bibnamefont
  {Ten~Tusscher}}\ and\ \bibinfo {author} {\bibfnamefont {A.~V.}\ \bibnamefont
  {Panfilov}},\ }\bibfield  {title} {\bibinfo {title} {Influence of diffuse
  fibrosis on wave propagation in human ventricular tissue},\ }\href@noop {}
  {\bibfield  {journal} {\bibinfo  {journal} {Europace}\ }\textbf {\bibinfo
  {volume} {9}},\ \bibinfo {pages} {vi38} (\bibinfo {year} {2007})}\BibitemShut
  {NoStop}%
\bibitem [{\citenamefont {Darling}\ and\ \citenamefont
  {Widrow}(2019)}]{darling2019eigenfunctions}%
  \BibitemOpen
  \bibfield  {author} {\bibinfo {author} {\bibfnamefont {K.}~\bibnamefont
  {Darling}}\ and\ \bibinfo {author} {\bibfnamefont {L.~M.}\ \bibnamefont
  {Widrow}},\ }\bibfield  {title} {\bibinfo {title} {Eigenfunctions of galactic
  phase space spirals from dynamic mode decomposition},\ }\href@noop {}
  {\bibfield  {journal} {\bibinfo  {journal} {Monthly Notices of the Royal
  Astronomical Society}\ }\textbf {\bibinfo {volume} {490}},\ \bibinfo {pages}
  {114} (\bibinfo {year} {2019})}\BibitemShut {NoStop}%
\bibitem [{\citenamefont {Barata}\ and\ \citenamefont
  {Hussein}(2012)}]{Barata2012}%
  \BibitemOpen
  \bibfield  {author} {\bibinfo {author} {\bibfnamefont {J.~C.~A.}\
  \bibnamefont {Barata}}\ and\ \bibinfo {author} {\bibfnamefont {M.~S.}\
  \bibnamefont {Hussein}},\ }\bibfield  {title} {\bibinfo {title} {The
  moore--penrose pseudoinverse: A tutorial review of the theory},\ }\href@noop
  {} {\bibfield  {journal} {\bibinfo  {journal} {Brazilian Journal of Physics}\
  }\textbf {\bibinfo {volume} {42}},\ \bibinfo {pages} {146} (\bibinfo {year}
  {2012})}\BibitemShut {NoStop}%
\bibitem [{\citenamefont {Penrose}(1955)}]{penrose1955generalized}%
  \BibitemOpen
  \bibfield  {author} {\bibinfo {author} {\bibfnamefont {R.}~\bibnamefont
  {Penrose}},\ }\bibfield  {title} {\bibinfo {title} {A generalized inverse for
  matrices},\ }in\ \href@noop {} {\emph {\bibinfo {booktitle} {Mathematical
  proceedings of the Cambridge philosophical society}}},\ Vol.~\bibinfo
  {volume} {51}\ (\bibinfo {organization} {Cambridge University Press},\
  \bibinfo {year} {1955})\ pp.\ \bibinfo {pages} {406--413}\BibitemShut
  {NoStop}%
\bibitem [{sup()}]{supmat}%
  \BibitemOpen
  \bibfield  {title} {\bibinfo {title} {Supplemental material},\ }\href@noop {}
  {\bibinfo  {journal} {Supplemental Material}\ }\BibitemShut {NoStop}%
\bibitem [{\citenamefont {Rowley}\ \emph {et~al.}(2009)\citenamefont {Rowley},
  \citenamefont {Mezi{\'c}}, \citenamefont {Bagheri}, \citenamefont
  {Schlatter},\ and\ \citenamefont {Henningson}}]{rowley2009spectral}%
  \BibitemOpen
\bibfield  {journal} {  }\bibfield  {author} {\bibinfo {author} {\bibfnamefont
  {C.~W.}\ \bibnamefont {Rowley}}, \bibinfo {author} {\bibfnamefont
  {I.}~\bibnamefont {Mezi{\'c}}}, \bibinfo {author} {\bibfnamefont
  {S.}~\bibnamefont {Bagheri}}, \bibinfo {author} {\bibfnamefont
  {P.}~\bibnamefont {Schlatter}},\ and\ \bibinfo {author} {\bibfnamefont
  {D.~S.}\ \bibnamefont {Henningson}},\ }\bibfield  {title} {\bibinfo {title}
  {Spectral analysis of nonlinear flows},\ }\href@noop {} {\bibfield  {journal}
  {\bibinfo  {journal} {Journal of fluid mechanics}\ }\textbf {\bibinfo
  {volume} {641}},\ \bibinfo {pages} {115} (\bibinfo {year}
  {2009})}\BibitemShut {NoStop}%
\bibitem [{\citenamefont {Brunton}\ \emph {et~al.}(2021)\citenamefont
  {Brunton}, \citenamefont {Budi{\v{s}}i{\'c}}, \citenamefont {Kaiser},\ and\
  \citenamefont {Kutz}}]{brunton2021modern}%
  \BibitemOpen
  \bibfield  {author} {\bibinfo {author} {\bibfnamefont {S.~L.}\ \bibnamefont
  {Brunton}}, \bibinfo {author} {\bibfnamefont {M.}~\bibnamefont
  {Budi{\v{s}}i{\'c}}}, \bibinfo {author} {\bibfnamefont {E.}~\bibnamefont
  {Kaiser}},\ and\ \bibinfo {author} {\bibfnamefont {J.~N.}\ \bibnamefont
  {Kutz}},\ }\bibfield  {title} {\bibinfo {title} {Modern koopman theory for
  dynamical systems},\ }\href@noop {} {\bibfield  {journal} {\bibinfo
  {journal} {arXiv preprint arXiv:2102.12086}\ } (\bibinfo {year}
  {2021})}\BibitemShut {NoStop}%
\bibitem [{\citenamefont {Fenton}\ \emph {et~al.}(2002)\citenamefont {Fenton},
  \citenamefont {Cherry}, \citenamefont {Hastings},\ and\ \citenamefont
  {Evans}}]{fenton2002multiple}%
  \BibitemOpen
  \bibfield  {author} {\bibinfo {author} {\bibfnamefont {F.~H.}\ \bibnamefont
  {Fenton}}, \bibinfo {author} {\bibfnamefont {E.~M.}\ \bibnamefont {Cherry}},
  \bibinfo {author} {\bibfnamefont {H.~M.}\ \bibnamefont {Hastings}},\ and\
  \bibinfo {author} {\bibfnamefont {S.~J.}\ \bibnamefont {Evans}},\ }\bibfield
  {title} {\bibinfo {title} {Multiple mechanisms of spiral wave breakup in a
  model of cardiac electrical activity},\ }\href@noop {} {\bibfield  {journal}
  {\bibinfo  {journal} {Chaos: An Interdisciplinary Journal of Nonlinear
  Science}\ }\textbf {\bibinfo {volume} {12}},\ \bibinfo {pages} {852}
  (\bibinfo {year} {2002})}\BibitemShut {NoStop}%
\bibitem [{\citenamefont {Laughner}\ \emph {et~al.}(2012)\citenamefont
  {Laughner}, \citenamefont {Ng}, \citenamefont {Sulkin}, \citenamefont
  {Arthur},\ and\ \citenamefont {Efimov}}]{laughner2012processing}%
  \BibitemOpen
  \bibfield  {author} {\bibinfo {author} {\bibfnamefont {J.~I.}\ \bibnamefont
  {Laughner}}, \bibinfo {author} {\bibfnamefont {F.~S.}\ \bibnamefont {Ng}},
  \bibinfo {author} {\bibfnamefont {M.~S.}\ \bibnamefont {Sulkin}}, \bibinfo
  {author} {\bibfnamefont {R.~M.}\ \bibnamefont {Arthur}},\ and\ \bibinfo
  {author} {\bibfnamefont {I.~R.}\ \bibnamefont {Efimov}},\ }\bibfield  {title}
  {\bibinfo {title} {Processing and analysis of cardiac optical mapping data
  obtained with potentiometric dyes},\ }\href@noop {} {\bibfield  {journal}
  {\bibinfo  {journal} {American Journal of Physiology-Heart and Circulatory
  Physiology}\ }\textbf {\bibinfo {volume} {303}},\ \bibinfo {pages} {H753}
  (\bibinfo {year} {2012})}\BibitemShut {NoStop}%
\bibitem [{\citenamefont {Miragoli}\ \emph {et~al.}(2007)\citenamefont
  {Miragoli}, \citenamefont {Salvarani},\ and\ \citenamefont
  {Rohr}}]{miragoli2007myofibroblasts}%
  \BibitemOpen
  \bibfield  {author} {\bibinfo {author} {\bibfnamefont {M.}~\bibnamefont
  {Miragoli}}, \bibinfo {author} {\bibfnamefont {N.}~\bibnamefont
  {Salvarani}},\ and\ \bibinfo {author} {\bibfnamefont {S.}~\bibnamefont
  {Rohr}},\ }\bibfield  {title} {\bibinfo {title} {Myofibroblasts induce
  ectopic activity in cardiac tissue},\ }\href@noop {} {\bibfield  {journal}
  {\bibinfo  {journal} {Circulation research}\ }\textbf {\bibinfo {volume}
  {101}},\ \bibinfo {pages} {755} (\bibinfo {year} {2007})}\BibitemShut
  {NoStop}%
\bibitem [{\citenamefont {Alonso}\ \emph {et~al.}(2016)\citenamefont {Alonso},
  \citenamefont {Dos~Santos},\ and\ \citenamefont
  {B{\"a}r}}]{alonso2016reentry}%
  \BibitemOpen
  \bibfield  {author} {\bibinfo {author} {\bibfnamefont {S.}~\bibnamefont
  {Alonso}}, \bibinfo {author} {\bibfnamefont {R.~W.}\ \bibnamefont
  {Dos~Santos}},\ and\ \bibinfo {author} {\bibfnamefont {M.}~\bibnamefont
  {B{\"a}r}},\ }\bibfield  {title} {\bibinfo {title} {Reentry and ectopic
  pacemakers emerge in a three-dimensional model for a slab of cardiac tissue
  with diffuse microfibrosis near the percolation threshold},\ }\href@noop {}
  {\bibfield  {journal} {\bibinfo  {journal} {PloS one}\ }\textbf {\bibinfo
  {volume} {11}},\ \bibinfo {pages} {e0166972} (\bibinfo {year}
  {2016})}\BibitemShut {NoStop}%
\bibitem [{\citenamefont {Maucher}\ and\ \citenamefont
  {Sutcliffe}(2017)}]{maucher2017length}%
  \BibitemOpen
  \bibfield  {author} {\bibinfo {author} {\bibfnamefont {F.}~\bibnamefont
  {Maucher}}\ and\ \bibinfo {author} {\bibfnamefont {P.}~\bibnamefont
  {Sutcliffe}},\ }\bibfield  {title} {\bibinfo {title} {Length of excitable
  knots},\ }\href@noop {} {\bibfield  {journal} {\bibinfo  {journal} {Physical
  Review E}\ }\textbf {\bibinfo {volume} {96}},\ \bibinfo {pages} {012218}
  (\bibinfo {year} {2017})}\BibitemShut {NoStop}%
\bibitem [{\citenamefont {Maucher}\ and\ \citenamefont
  {Sutcliffe}(2019)}]{maucher2019dynamics}%
  \BibitemOpen
  \bibfield  {author} {\bibinfo {author} {\bibfnamefont {F.}~\bibnamefont
  {Maucher}}\ and\ \bibinfo {author} {\bibfnamefont {P.}~\bibnamefont
  {Sutcliffe}},\ }\bibfield  {title} {\bibinfo {title} {Dynamics of linked
  filaments in excitable media},\ }\href@noop {} {\bibfield  {journal}
  {\bibinfo  {journal} {Nonlinearity}\ }\textbf {\bibinfo {volume} {32}},\
  \bibinfo {pages} {942} (\bibinfo {year} {2019})}\BibitemShut {NoStop}%
\bibitem [{\citenamefont {Weingard}(2017)}]{weingard2017scroll}%
  \BibitemOpen
  \bibfield  {author} {\bibinfo {author} {\bibfnamefont {D.}~\bibnamefont
  {Weingard}},\ }\emph {\bibinfo {title} {Scroll Waves and How They Interact
  with Non-Reactive Spheres, Tori, and Knots}},\ \href@noop {} {Ph.D. thesis},\
  \bibinfo  {school} {The Florida State University} (\bibinfo {year}
  {2017})\BibitemShut {NoStop}%
\bibitem [{\citenamefont {Rajany}\ \emph {et~al.}(2021)\citenamefont {Rajany},
  \citenamefont {Majumder}, \citenamefont {Nayak},\ and\ \citenamefont
  {Pandit}}]{rajany2021effects}%
  \BibitemOpen
  \bibfield  {author} {\bibinfo {author} {\bibfnamefont {K.~V.}\ \bibnamefont
  {Rajany}}, \bibinfo {author} {\bibfnamefont {R.}~\bibnamefont {Majumder}},
  \bibinfo {author} {\bibfnamefont {A.~R.}\ \bibnamefont {Nayak}},\ and\
  \bibinfo {author} {\bibfnamefont {R.}~\bibnamefont {Pandit}},\ }\bibfield
  {title} {\bibinfo {title} {The effects of inhomogeneities on scroll-wave
  dynamics in an anatomically realistic mathematical model for canine
  ventricular tissue},\ }\href@noop {} {\bibfield  {journal} {\bibinfo
  {journal} {Physics Open}\ }\textbf {\bibinfo {volume} {9}},\ \bibinfo {pages}
  {100090} (\bibinfo {year} {2021})}\BibitemShut {NoStop}%
\end{thebibliography}%

\end{document}


\title{Supplemental Material for Spiral-wave dynamics 
in excitable media: Insights from dynamic mode decomposition.}

\author{Mahesh Kumar Mulimani}
\email{maheshk@iisc.ac.in ;}
\affiliation{Centre for Condensed Matter Theory, Department of Physics, Indian Institute of Science, Bangalore 560012, India.}
\author{Soling Zimik}%
\email{solyzk@gmail.com ;}
\affiliation{Computational Biology Group, Institute of Mathematical Sciences, CIT Campus, Tharamani, Chennai, 600113, India.}
\author{Jaya Kumar Alageshan}
\email{jayak@iisc.ac.in ;}
\affiliation{Centre for Condensed Matter Theory, Department of Physics, Indian Institute of Science, Bangalore 560012, India.}
\author{Rahul Pandit}
 \email{rahul@iisc.ac.in}
 \affiliation{Centre for Condensed Matter Theory, Department of Physics, Indian Institute of Science, Bangalore 560012, India.}

\maketitle

\begin{figure*}
   \centering
   \includegraphics[scale=0.25]{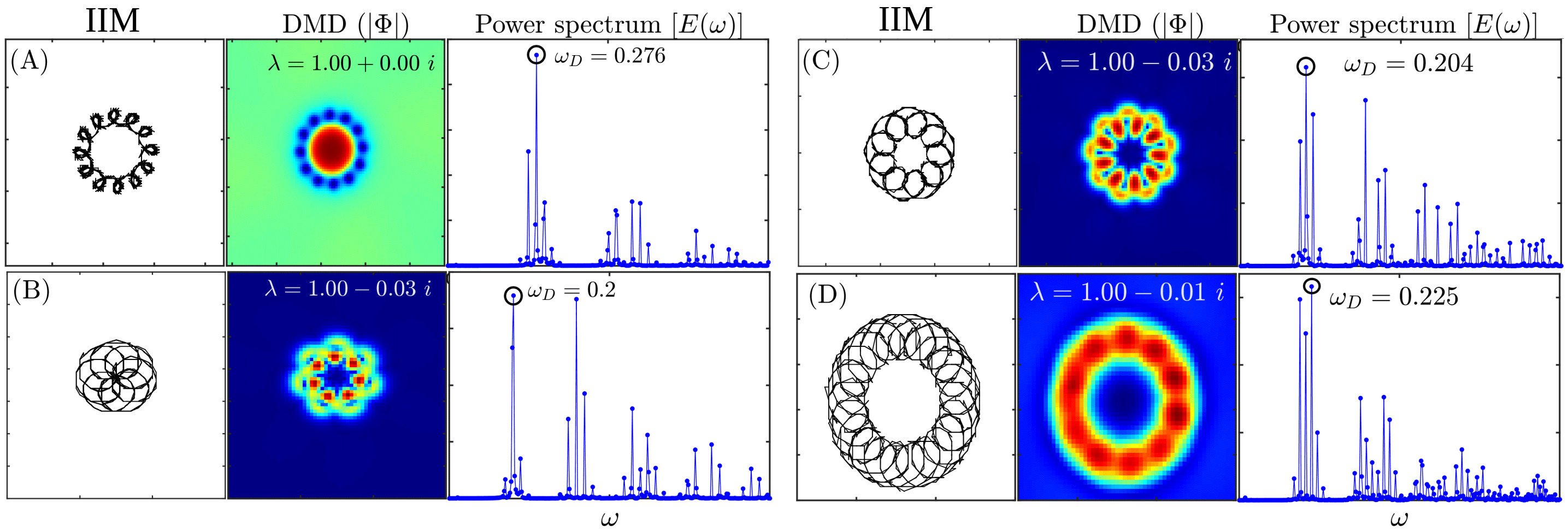}
    \caption{Illustrative plots, for the Barkley model [Eqs.~(1)
and (2) in the main paper], showing four different types of tip-trajectories
(TTs) [apart from the ones shown in Fig. $2$ in the 
main manuscript]. In each one of the panels (A)-(D), the sub-figures in the left
columns show the TTs obtained from the IIM (see text), those in the
middle columns show, via pseudocolor plots, of the modulus of
a DMD eigenmode $\Phi$ (see text), with eigenvalue $\lambda$,
that contains an imprint of the TT, and those in the right column show
the power spectrum, obtained from the time series of $u$ (cf.
Fig. $1$ in the main paper). 
The TTs in (A)-(D) display aperiodic motions with more than one fundamental i
frequency.}
    \label{fig:s1}
\end{figure*}

In this Supplemental Material we provide the following:
\begin{itemize}
\item The DMD algorithm that is used to calculate the DMD eigenmodes $\Phi_i$. 
\item In Fig.~\ref{fig:s1} we show the different types of tip-trajectories 
that we observe in the Barkley Model [Eqs. (1) and (2) in the main paper]
apart from the ones that we have shown in the main manuscript.
\item In Table.~\ref{Table:S1} we list the parameters, $a$, $b$, and $\epsilon$  that we use in our studies of the Barkley Model [Eqs. (1) and (2) in the main paper].
\item Movie-M1: the spatiotemporal evolution of the spiral wave $u$,
DMD-reconstructed wave $u^P$, and the error $u-u^P$ (see Fig. $7$ in the
main paper).
\end{itemize}

\section{DMD algorithm}
\begin {itemize}
\item[1.] Define the two data matrices $X_1$ and $X_2$ (see Eq. (4) 
and Fig. $1$ in the main paper) as follows:
\begin{eqnarray}
    {X_1}=\begin{bmatrix}
        | &  & |\\
        x_0 & ... & x_{m-1}\\
        |& &|
        \end{bmatrix};
    {X_2}=\begin{bmatrix}
        | &  & |\\
        x_1 & ... & x_{m}\\
        |& &|
        \end{bmatrix}.
\end{eqnarray}
\begin{figure}
\includegraphics[scale=0.35]{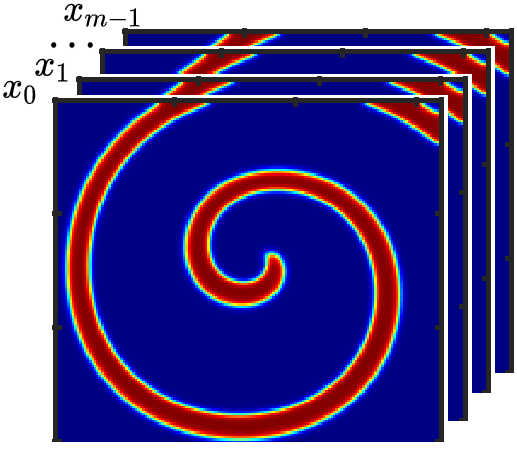}
	\caption{Pseudocolor plots of $u$ (cf. Fig. $1$ in the main paper) 
showing the $m$ spiral waves, at different instants, which we use to
construct the data matrix $X_1$.}   
    \label{fig:figs0}
\end{figure} 
\item [2.] The linear operator $\mathcal{L}$ is 
\begin{eqnarray}
    \mathcal{L}=X_2 X_1^{\dagger};
\end{eqnarray}
cf., Eq. (5) of the main paper.

\item [3.] We perform singular value decomposition (SVD) on $X1$ as follows:
\begin{eqnarray}
    X_1=U \Sigma V^{\dagger}
\end{eqnarray}

\item [4.] The singular values $S$, which contain information about the 
low-rank structure in the system, are
\begin{equation}
S = \sqrt{\Sigma_{ii}} .
\end{equation}
\item [5.]  The first $r$ left- and right-singular matrices $U_r$ and $V_r$, 
respectively, can approximate $X_1$: 
\begin{equation}
{X_1} \approx U_r \Sigma_r V_{r}^{T} .
\end{equation}
\item [6.] We transform the matrix $\mathcal{L}$ ($n \times n$) into its
low-rank structure $ \mathcal{L}_r$ ($r \times r$) through the 
similarity transformation
\begin{eqnarray}
 \mathcal{L}_r = U^T_r A U_r . \\
  \mathcal{L}_r = U_r^T X_2 V_{r} \Sigma_r^{-1} .
\end{eqnarray}
\item [7.] We then find the eigenvalues $\lambda_i$ and the eigenvectors 
$W_i$ of $\mathcal{L}_r$:
\begin{eqnarray}
\mathcal{L}_r \ W_i = W_i \lambda_i .  
\end{eqnarray}
\item [8.] The eigenvector $W_i$ is now transformed into the 
higher-dimensional space ($n$), to obtain the eigenvector: 
\begin{eqnarray}
 \Phi_i = U_r^{T} {X_2} V_r \Sigma_r^{-1} W_i . 
\end{eqnarray} 
\end{itemize}


\begin{table*}
\resizebox{13cm}{!}{
\begin{tabular}{|l|l|l|l|}
\hline
\ $a$ & \ \ $b$ \ & \ \ $\epsilon$ \  \ &  \ Nature of tip-trajectory  \ \\ 
\hline
\hline
 0.52 \ & \ 0.05  \ & \ 0.017 \ & \ Inward Petals \\ \hline

 0.503 \ & \ 0.05  \ & \ 0.017 \ & \ Rosette Pattern \\\hline

 0.428 \ & \ 0.05  \ & \ 0.016 \ & \ Circle \\\hline

  0.58 \ & \ 0.05  \ & \ 0.016 \ & \ Outward Petals \\\hline

 0.48 \ & \ 0.015 \  & \ 0.02 \ & \ Rectangular Pattern \\\hline

\hline
\end{tabular}}
\caption{The parameter values we use for the Barkley Model [Eqs. (1) and (2) 
in the main paper] that we use to generate different tip-trajectory patterns.}
\label{Table:S1}
\end{table*}

\bibliography{references}